\documentclass[%
 reprint,
superscriptaddress,
showkeys,
showpacs,
nofootinbib,
 amsmath,amssymb,
 aps,
]{revtex4-1}
\usepackage{graphicx,subfigure}
\usepackage{dcolumn}
\usepackage{bm}
\usepackage{wrapfig}

\usepackage{mathtools}
\begin{document}

\preprint{APS/123-QED}

\title[Gravitational wave generation by interaction of high power lasers with matter]{Gravitational wave generation by interaction of high power lasers with matter. \\
Part II: Ablation and Piston models}

\author{Hedvika Kadlecov\'{a}}
 \email{Hedvika.Kadlecova@eli-beams.eu}
 \affiliation{Institute of Physics of the ASCR, ELI--Beamlines project, Na Slovance 2, 18221, Prague, Czech republic}
 \author{Ond\v{r}ej Klimo}
 \affiliation{Institute of Physics of the ASCR, ELI--Beamlines project, Na Slovance 2, 18221, Prague, Czech republic}
 \affiliation{FNSPE, Czech Technical University in Prague, 11519 Prague, Czech Republic}
 \author{Stefan Weber}
 \affiliation{Institute of Physics of the ASCR, ELI--Beamlines project, Na Slovance 2, 18221, Prague, Czech republic}
 \author{Georg Korn}
 \affiliation{Institute of Physics of the ASCR, ELI--Beamlines project, Na Slovance 2, 18221, Prague, Czech republic}

\date{\today}

\begin{abstract}
We analyze theoretical models of gravitational waves generation in the interaction of high intensity laser with matter. We analyse the generated gravitational waves in linear approximation of gravitational theory. We derive the analytical formulas and estimates for the metric perturbations and the radiated power of generated gravitational waves. Furthermore we investigate the characteristics of polarization and the behaviour of test particles in the presence of gravitational wave which will be important for the detection. 

\begin{description}
\item[PACS numbers]\pacs{52.38.-r, 04.30.Db, 52.27.Ey, 52.38.Kd}
\end{description}
\end{abstract}

\pacs{52.38.-r, 04.30.Db, 52.27.Ey, 52.38.Kd}
\keywords{gravitational waves, laser--plasma interaction, generation of gravitational waves, experiment}
\maketitle

\section{\label{sec:number1}Introduction}
The main purpose of the second part of the paper is to properly analyze other two generation models of high frequency gravitational waves (HFGW) in the interaction of high power laser pulse with a medium, the ablation (rarefaction) \cite{Fabbro1984}  and piston \cite{Naumova2009} models. These models were suggested in \cite{grossmannMeet2009l, izestELINP2014}. The theory and the basic information about the models was reviewed in the part I \cite{Kadlecova2015} where we investigated the shock wave model in detail. Therefore we will move faster in this second part and will concentrate on new results for the ablation and piston models.

The paper is organized as follows. In Section \ref{sec:number2}, we derive and analyze the analytical formulae for the perturbations and the luminosity of the gravitational radiation. We present the estimations for the experiment and measurement for the specific data for ablation model.

In Section \ref{sec:number3}, we concentrate on the piston model and provide the analytical formulae for perturbation, the luminosity and estimations for an experiment. 

In Section \ref{sec:number4} we derive and analyze the polarization properties of the gravitational radiation and the different radiative properties with dependence on the orientation of the wave vector in the assumed ablation and piston model.

The Section \ref{sec:number5} we concentrate on derivation and analysis of the behaviour of the test particles in the field of passing gravitational waves in both models, ablation and piston one.
 
The main results are summarized in the concluding Section \ref{sec:number6}.

\section{\label{sec:number2}The derivation of gravitational wave characteristics for ablation model}
The calculations are made in linear approximation to full gravity theory \cite{MaggioreBook,BicakBook, MTWBook} up to quadrupole moment in the multipole expansion, for details in theory see \cite{Kadlecova2015}.

In the configuration pictured in Fig. 1 in \cite{Kadlecova2015}, the laser is interacting with a planar thick foil with more than 100 $\mu m$ thickness. The material is accelerated in the ablation zone and in the shock front. The points on the axis $z_{a}$ and $z_{s}$ indicate the areas where the gravitational waves start to be generated. These two possibilities are divided into two separate models \cite{Fabbro1984}, the shock wave model and the ablation zone generation model. In the experiment, the two models are put together since each model represents one faze of the same experiment and therefore the radiation could be measured simultaneously.


In the following text we are going to investigate the ablation model in detail.

\subsection{The ablation zone generation model}
In this case the gravitational radiation is produced in the ablatation zone with starting point $z_{r}$. The density profile  for this model is visible in the Fig.~\ref{fig:wide3}. The expressions will be very similar to ones for the shock wave model therefore we will proceed in a shorter way.

\subsubsection{\label{eq:lim1}The limitations of the theory}
Let see whether the low velocity limit Eq.~(7) in \cite{Kadlecova2015} is satisfied for ablation model. The linear size (diameter) of the source (the focus size) is $d= 1 {\rm mm}= 10^{-3}\,{\text m}$  and the reduced generated wavelength is $\lambdabar=4.7746\times 10^{-2} $~m for the gravitational wave length $\lambda_{g}=0.3$ m, which is the same as for shock wave model \cite{Kadlecova2015}.
The comparison Eq.~(7) in \cite{Kadlecova2015} results into
\begin{equation}
0.021 \ll 1.\label{eq:compLambdaA}
\end{equation}
The low velocity condition is still satisfied for the ablation wave experiment, while we have obtained the condition for the size of the target to satisfy the low velocity condition.
We can generalize the estimation with the fact that 
\begin{align}
\lambdabar=\frac{1}{2\pi}\tau c,\label{eq:LimitEstimaceGen}
\end{align}
where $\tau$ is the duration of the pulse and $c$ is the speed of light, then we can rewrite this condition as
\begin{align}
d \ll \frac{1}{2\pi}\tau c,\label{eq:destimationgeneral}
\end{align}
which could be useful in general setup of the experiment according to the duration of the pulse.

\subsubsection{Set up of the experiment}
This section is devoted to the derivation of fully analytical formulae of the luminosity ${\mathcal{L}}_{GW}$ and the perturbation of the metric $h_{GW}$ for the shock wave model in Section \ref{sec:number3} using the linerized gravity theory from Section \ref{sec:number2}. The results are new, as well as the results in the following sections about polarization and behaviour of test particles in the gravitational field of gravitational wave.

\begin{figure}[h!]
\centering
\includegraphics[width=0.45\textwidth]{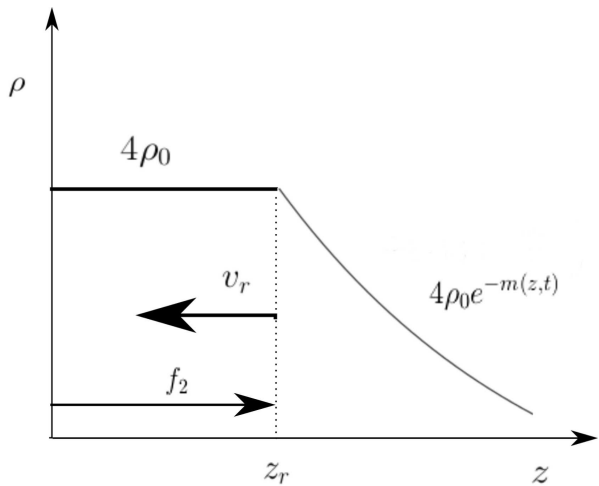}
\caption{\label{fig:wide3}The representation of density profile for the ablation zone model.}
\end{figure} 

The set up of the geometry of the experiment is similar to shock wave model.
We assume the rectangular shape of the foil with parameters, a, b, l, and we choose the orthogonal coordinate system x, y, z. The parameter $l$ is the thickness of the foil in the $z$ direction. The distance of the laser and the detection desk/point is $z_{L}$, for full set up see Fig.~3 in \cite{Kadlecova2015}. We assume the whole process happends in the box of rectangular shape with parameters $a, b, z_{L}$ for simplicity. The start of the coordinate system corresponds with the position where the detector would be possibly positioned. The moving point where the density of the beam changes will be denoted as $z_{r}$ with a form 
\begin{align}
z_{r}(t)= -v_{r}t + d, \label{eq:zr}
\end{align}
where the velocity is defined as
\begin{equation} 
 v_{r}(t)\approx c_{r}=\sqrt{\frac{P_{r}}{4\rho_{0}}},\label{eq:vrdef}
\end{equation}
where $P_{r}$ is ablation pressure and $\rho_{0}$ is material density. 
We assume that for $t=0$, $z_{s}(0)=d$, therefore the constant in (\ref{eq:zr}) is $d=f_{2}$ according the the Fig.~\ref{fig:wide3}.

In the following, we will calculate everything with general function $z_{r}(t)$ and then we will substitute the explicit function (\ref{eq:zr}) at convenient places. General expressions might be useful for other forms of $z_{r}(t)$. At this point in time, we are not aware of better ansatz for this function.
 
The basic input for the calculation is the density profile from Fig.~\ref{fig:wide3}. The step function for the density profile can be written as
\begin{equation}
\rho(t,\bf{x})=
\begin{cases}
\;4\rho_{0} &\quad if \quad z < z_{r},\label{eq:densityAblation}\\
\;4\rho_{0}e^{-m(z,t)} &\quad if \quad z > z_{r},
\end{cases}
\end{equation}
where we denote $m(z,t)$ as
\begin{equation}
m(z,t)=-\frac{z-z_{r}}{z_{r}}\label{eq:m}.
\end{equation}
The density does not satisfy the mass conservation law because we integrate the mass moment to the finite value $z_{L}$ instead of the $\infty$ value. This property of the ablation model has its concequences in obtaining artificial gravitational waves in the direction of the laser propagation, which will be discussed later in the paper. Such a property of a model was also observed in \cite{Mora}.

The first step in the calculation is the mass moment derivation.

\subsubsection{The mass moment}
The values for integration of the density (\ref{eq:densityAblation}) in Eq.~(11) in \cite{Kadlecova2015} are $x\in <0,a)$, $y\in <0,b)$ and $z\in<0,z_{L}>$ which splits into $<0,z_{r})$ and $(z_{r},z_{L}>$. In other words, we integrate over the box in the Fig.~3 in \cite{Kadlecova2015}. 
We denote 
\begin{align}
a_{I}\equiv m(z_{r},t)=0,&\quad b_{I}\equiv m(z_{L},t)=-\frac{z_{L}-z_{r}}{z_{r}},\label{eq:mztf}
\end{align}
and when $z=0$ the function $m(0,t)=1$ for every $t$.

The mass moment Eq.~(11) in \cite{Kadlecova2015} is listed in Appendix (\ref{eq:componentsMassMomentDiagApp}) and (\ref{eq:componentsMassMomentOffApp}) where we used (\ref{eq:mztf}), then the diagonal components then read
\begin{align}
M_{xx}&=\frac{4}{3}Sa^2\rho_{0}z_{r}e^{-b_{I}},\quad
M_{yy}=\frac{4}{3}Sb^2\rho_{0}z_{r}e^{-b_{I}},\nonumber\\ 
M_{zz}&=4S\rho_{0}z_{s}^3\left(-\frac{2}{3}+(b^2_{I}+1)e^{-b_{I}}\right),\label{eq:componentsMassMomentDiag}
\end{align}
and non--diagonal components $M_{xy}, M_{yz}, M_{xz}$,
\begin{align}
M_{xy}&=S^2\rho_{0}z_{r}e^{-b_{I}},\;
M_{yz}=2Sb\rho_{0}z_{r}^2\left(\frac{1}{2}+b_{I}e^{-b_{I}}\right),\nonumber\\
M_{xz}&=2Sa\rho_{0}z_{r}^2\left(\frac{1}{2}+b_{I}e^{-b_{I}}\right).\label{eq:componentsMassMomentOff}
\end{align}

\subsubsection{The quadrupole moment}
The next step is the calculation of the quadrupole moment Eq.~(10) in \cite{Kadlecova2015}. The non--diagonal components $I_{xy}, I_{yz}, I_{xz}$ are
\begin{align}
I_{xy}&=M_{xy},\quad I_{yz}=M_{yz},\quad I_{xz}=M_{xz}.\label{eq:componentsQuadrMomentOff}
\end{align}
The diagonal components $I_{ii}=M_{ii}-\frac{1}{3}Tr M$ read
\begin{align}
I_{xx}&=\frac{4S\rho_{0}}{3}z_{r}\left\{z_{r}^2(\frac{2}{3}-(b^2_{I}+1)e^{-b_{I}})+\frac{(2a^2-b^2)}{3}e^{-b_{I}}\right\},\nonumber\\
I_{yy}&=\frac{4S\rho_{0}}{3}z_{r}\left\{z_{r}^2(\frac{2}{3}-(b^2_{I}+1)e^{-b_{I}})+\frac{(2b^2-a^2)}{3}e^{-b_{I}}\right\},\nonumber\\
I_{zz}&=\frac{4S\rho_{0}}{3}z_{r}\left\{4z_{r}^2((b^2_{I}+1)e^{-b_{I}}-\frac{2}{3})-\frac{(a^2+b^2)}{3}e^{-b_{I}}\right\}.\label{eq:componentsQuadDiag}
\end{align}
Similarly to the shock wave model, the diagonal components of quadrupole moment show cubic dependence on the function $z_{r}$ and are missing quadratic term. The non-diagonal components $I_{yz}$ and $I_{xz}$ are missing the linear dependence on $z_{r}$.
The trace $Tr M_{ii}$ reads
\begin{align}
Tr M_{ii}&=\frac{4}{3}S\rho_{0}z_{r}\left[(a^2+b^2)e^{-b_{I}}+3z_{r}^2(-\frac{2}{3}+(1+b^2_{I})e^{-b_{I}})\right].
\end{align}

When we substitute the function $z_{r}(t)$ into $I_{zz}$ component we will get the time dependency as
\begin{align}
I_{zz}&=\frac{4S\rho_{0}}{3}\left\{4(-v^3_{r}t^3+3v_{r}^2t^2f_{2}-3v_{r}tf^2_{2}+f_{2}^2)\times \right.\nonumber\\
&\left.(-\frac{2}{3}+(2\frac{z_{L}}{z_{r}}-1)e^{-b_{I}})-(-v_{r}t+f_{2})\frac{(a^2+b^2)}{3}e^{-b_{I}}\right\}.\label{eq:IzzWithSubst}
\end{align}
The quadrupole moment in the $zz$ direction is given by a cubic polynomial in $t$ variable as in the shock model \cite{Kadlecova2015}. The most dominant term is then the cubic term with a new term $e^{-b_{I}}$ which behaves as $e^{-1}$ when $t\rightarrow 0$ and creates dumping as time progresses. The other terms are new, the quadratic, linear and constant terms. The geometry of the setup influences the quadrupole moment from the quadratic term and lower.

\subsubsection{The analytical form of perturbation and luminosity}
Now, we calculate the components of the perturbation tensor according to Eq.~(9) in \cite{Kadlecova2015} without projector $\Lambda_{ij,kl}$. In other words, we got the components of the perturbation tenzor in general form, the components read
\begin{align}
h_{xx}&=\frac{8G}{3rc^4}S\rho_{0}\left\{\frac{2a^2-b^2}{3}\frac{z^2_{L}}{z^3_{r}}e^{-b_{I}}-(z^3_{r}D)^{\ddot{}}\right\},\nonumber\\
h_{yy}&=\frac{8G}{3rc^4}S\rho_{0}\left\{\frac{2a^2-b^2}{3}\frac{z^2_{L}}{z^3_{r}}e^{-b_{I}}-(z^3_{r}D)^{\ddot{}}\right\},\label{eq:DiagAbl}\\
h_{zz}&=\frac{8G}{3rc^4}S\rho_{0}\left\{-\frac{a^2+b^2}{3}\frac{z^2_{L}}{z^3_{r}}e^{-b_{I}}+4(z^3_{r}D)^{\ddot{}}\right\},\nonumber
\end{align}
and the non-diagonal terms are
\begin{align}
h_{xy}&=-\frac{2G}{rc^4}S^2\rho_{0}z^2_{L}\frac{e^{-b_{I}}}{z^3_{r}},\nonumber\\
h_{xz}=\frac{4G}{rc^4}Sa\rho_{0}&\left\{(z^2_{r})^{\dot{}}(\frac{1}{2}+b_{I}e^{-b_{I}})+e^{-b_{I}}\frac{z^2_{L}}{z^3_{r}}(2(z^2_{r})^{\dot{}}-(z_{r}+z_{L}))\right\},\label{eq:NonAbl}\\
h_{yz}=\frac{4G}{rc^4}Sb\rho_{0}&\left\{(z^2_{r})^{\dot{}}(\frac{1}{2}+b_{I}e^{-b_{I}})+e^{-b_{I}}\frac{z^2_{L}}{z^3_{r}}(2(z^2_{r})^{\dot{}}-(z_{r}+z_{L}))\right\},\nonumber
\end{align}
where we have used 
\begin{equation}
D=-\frac{2}{3}+e^{-b_{I}}(b^2_{I}+1),\label{eq:D}
\end{equation}
and conveniently $\ddot{z}_{r}=0$ for substitution (\ref{eq:zr}) to simplify the expressions. We are not going to list all the derivatives in Appendix for this model because of the complexity of expressions.

Contrary to the shock wave model calculations, all components  of $h_{ij}$ are time dependent components of the tenzor thanks to functions  $a_{I}$ and $b_{I}$. Just in the diagonal components the first term vanishes for 
\begin{equation}
f_{2}=v_{r}t, \label{eq:vanishVr}
\end{equation}
which is the position of the detector.

We will investigate the component $zz$ of perturbation because it is the most complex component in the direction of motion of the experiment, the components $h_{xx}$ and $h_{yy}$ has similar terms in their expression and therefore for the purposes of estimation and functional dependence it is enought to investigate just $zz$ component.

First, we investigate the component of perturbation $h^{GW}_{zz}$ which can be rewritten as 
\begin{align}
h_{zz}&=\frac{8G}{3rc^4}S\rho_{0}\left(24(\dot{z}_{r})^2\left[-z_{r}(\frac{2}{3}+e^{-b_{I}})+\frac{3}{2}z_{L}e^{-b_{I}}\right]\right.\nonumber\\
+&\left.24z_{L}\dot{z_{r}}e^{-b_{I}}(1-\frac{z_{L}}{z_{r}})+4e^{-b_{I}}(4z_{L}-3z_{r})\frac{z^2_{L}}{z^2_{r}}\right.\label{eq:hrewritten}\\
&\left.-\frac{(a^2+b^2)}{3}e^{-b_{I}}\frac{z^2_{L}}{z^3_{r}}\right).\nonumber
\end{align}
For the purposes of an estimation we will evaluate just the first term of (\ref{eq:hrewritten}) which is linear in $z_{r}$ and most dominant. The second term behaves as $O(\frac{e^{-b_{I}}}{z_{r}})$, the third as $O(\frac{e^{-b_{I}}}{z^2_{r}})$ and the fourth as $O(\frac{e^{-b_{I}}}{z^3_{r}})$ which in limit $t\rightarrow \infty$ approach zero. According to the fourth term the parameters of the foil then contribute in the small way to the value of perturbation.
 
The expression (\ref{eq:hrewritten}) becomes using 
(\ref{eq:zr}), (\ref{eq:vrdef}), 
\begin{align}
h_{zz}&=\frac{64G}{rc^4}\left(v^3_{r}t(\frac{2}{3}+e^{-b_{I}})-v^{2}_{r}\left[f_{2}(\frac{3}{2}+e^{-b_{I}})-\frac{3}{2}z_{L}e^{-b_{I}}\right]\right).\label{eq:rewritten}
\end{align}

The previous expression can be rewritten even further using 
(\ref{eq:vrdef}) and (\ref{eq:Pressure}) as
\begin{align}
h_{zz}&=\frac{64G}{rc^4}\left(\frac{1}{6}\left(\frac{R_{t}}{\rho_{0}}\right)^{1/2}E_{L}(\frac{2}{3}+e^{-b_{I}})\right.\\
-&\left.\frac{SR_{t}^{1/3}I^{2/3}_{L}}{4}\left[f_{2}(\frac{2}{3}+e^{-b_{I}})-\frac{3}{2}z_{L}e^{-b_{I}}\right]\right).\label{eq:rewrittenAgain}
\end{align}
where we used the pressure and the energy of the laser, 
\begin{equation}
P_{L}=SI_{L},\quad E_{L}=SI_{L}t.\label{eq:pressureAndenergy}
\end{equation}
When we compare this final formula with one for shock wave model \cite{Kadlecova2015} we observe that the perturbation is more general in terms with $e^{-b_{I}}$. This is a natural consequence of the more general density ansatz (\ref{eq:densityAblation}) when compared with one for shock wave model. Thanks to the ansatz the constant $z_{L}$ appears in the final expression. The value of the perturbation decreases with the distance as $1/r$ and will be zero in the infinity. We have obtained additional time dependent terms which contribite to the first term in the brackets.

We use more general expression for $P_{r}$ and  $I_{L}$ \cite{AtzeniBook} which will allow us to have control over more parameters than the formulae suggested in \cite{izestELINP2014, grossmannMeet2009l},
\begin{equation}
P_{r}=R_{t}^{1/3}I_{L}^{2/3}, \label{eq:Pressure}
\end{equation}
and $R_{t}$ denotes the target 'density' as $R_{t}=\frac{1}{2}\frac{A}{Z}m_{p}n_{c}$,
and $n_{c}$ is the critical density defined as $n_{c}=\frac{\epsilon_{0}m_{e}}{e^2}\frac{(2\pi c)^2}{\lambda_{L}^2}$,
where $\epsilon_{0}$ is vacuum permitivity of vacuum, $m_{e}$ is the rest mass of the electron, $e$ is the charge of electron and $\lambda_{L}$ is the wave length of the laser. All of the parameters in $n_{c}$ are constants except the laser wavelenght $\lambda_{L}$ which is constant given by the specific experiment.

The luminosity Eq.~(12) can be rewritten as Eq.~(27) in \cite{Kadlecova2015}.
After substituting the quadrupole moment components into  Eq.~(27) in \cite{Kadlecova2015}, we get general expression as
\begin{align}
&\mathcal{L}_{\text{quad}}=\frac{G}{5c^5}S^2\rho^2_{0}\left\{\frac{16}{9}\left\{18[(z^3_{r}(-\frac{2}{3}+e^{-b_{I}}(b^2_{I}+1)))^{\dddot{}}]^2\right.\right.\nonumber\\
-&\left.\left.\frac{10}{3}[{z}_{r}e^{-b_{I}}]^{\dddot{}}[z^3_{r}(-\frac{2}{3}+e^{-b_{I}}(b^2_{I}+1))]^{\dddot{}}(a^2+b^2)\right.\right.\nonumber\\
 +&\left.\left.\frac{1}{9}[({z}_{r}e^{-b_{I}})^{\dddot{}}]^2[(a^2+b^2)^2+(2a^2-b^2)^2+(2b^2-a^2)^2+\frac{81}{8}S^2]\right\}\right.\nonumber\\
 +&\left.8(a^2+b^2)[(z^2_{r}(\frac{1}{2}+b_{I}e^{-b_{I}}))^{\dddot{}}]^2\right\}. \label{eq:LumRes2}
\end{align}
We observe that the expression is in fact generalized luminosity for shock wave model \cite{Kadlecova2015} with terms with $b_{I}$ as in previous results. Contrary to result for shock  wave model the result it time dependent. In order to obtain the most dominant contribution we neglect the higher derivatives of terms with $b_{I}$ because the higher the derivative of such terms the higher the power of $z_{r}$ in 
denominator and lower contribution. Then we obtain
\begin{align}
&\mathcal{L}_{\text{quad}}=\frac{G}{5c^5}S^2\rho^2_{0}\left\{\frac{16}{9}\left\{18[(z^3_{r})^{\dddot{}}(-\frac{2}{3}+e^{-b_{I}}(b^2_{I}+1))]^2\right.\right.\nonumber\\
-&\left.\left.\frac{10}{3}[({z}_{r})^{\dddot{}}e^{-b_{I}}][(z^3_{r})^{\dddot{}}(-\frac{2}{3}+e^{-b_{I}}(b^2_{I}+1))](a^2+b^2)\right.\right.\nonumber\\
 +&\left.\left.\frac{1}{9}[({z}_{r})^{\dddot{}}e^{-b_{I}})]^2[(a^2+b^2)^2+(2a^2-b^2)^2+(2b^2-a^2)^2+\frac{81}{8}S^2]\right\}\right.\nonumber\\
 +&\left.8(a^2+b^2)[(z^2_{r})^{\dddot{}}(\frac{1}{2}+b_{I}e^{-b_{I}}))]^2\right\}. \label{eq:LumRes3}
\end{align}
which further simplifies to 
\begin{align}
\mathcal{L}_{\text{quad}}=&\frac{1152G}{5c^5}S^2\rho_{0}^2v^6_{s}[-\frac{2}{3}+e^{-b_{I}}(b^2_{I}+1)]^2. \label{eq:LumRes4}
\end{align}

Finally, we will use the explicit expression for the velocity $v_{s}$ via (\ref{eq:vrdef}) and (\ref{eq:Pressure}), we will obtain the final expression for luminosity of gravitational radiation,
\begin{align}
\mathcal{L}_{\text{quad}}=&\frac{9G}{10c^5}\frac{R_{t}P^2_{L}}{\rho_{0}^2}[-\frac{2}{3}+e^{-b_{I}}(b^2_{I}+1)]^2, \label{eq:LumResFinAb}
\end{align}
where the first term in the brackets is constant, second one is $O(\frac{e^{-b_{I}}}{z_{r}})$ and third one $O(\frac{e^{-b_{I}}}{z^2_{r}})$. The terms with $b_{I}$ are corrections to the most dominant constant term.
The luminosity then depends on the power of the laser, the density of the material and the laser wavelength. The result generalizes \cite{grossmannMeet2009l,izestELINP2014} in the dependency on the laser wavelength and correction terms with $b_{I}$ and constant $R_{t}$. The numerical factor in front of the fraction for estimation will be presented in the next subsection.

Interestingly, the quadrupole moment using (\ref{eq:pressureAndenergy}),
\begin{align}
I_{zz}(t)&=\frac{4S\rho_{0}}{3}\left\{\left[-\frac{R_{t}^{1/2}E_{L}}{\rho_{0}^{3/2}}t^2+\frac{R_{t}^{1/3}I_{L}^{2/3}f_{2}}{\rho_{0}}t\right.\right.\nonumber\\
-&\left.\left.\frac{6R_{t}^{1/6}I_{L}^{1/3}f_{2}^2}{\sqrt{\rho_{0}}}+f^2_{2}\right]\times\left(-\frac{2}{3}+(2\frac{z_{L}}{z_{r}}-1)e^{-b_{I}}\right)\right.\nonumber\\
&-\left.\frac{(a^2+b^2)}{3}e^{-b_{I}}(-\frac{6(R_{t}^{1/2}I_{L})^{1/3}}{\sqrt{\rho_{0}}}t+f_{2})\right\}.\label{eq:IzzWithVsP}
\end{align}
has similar form as for the shock wave model \cite{Kadlecova2015} generalized with terms $b_{I}$.  

In this subsection, we have derived explicit expressions for perturbation component $h^{GW}_{zz}$ and $\mathcal{L}_{\text{quad}}$ which generalize previosly published results with additional time dependent terms with function $b_{I}$ and constant $R_{t}$.

\subsubsection{\label{subsec:estimate}The estimations for the $h_{\mu\nu}$ and $\mathcal{L}_{\text{quad}}$ for real experiment}
We will evaluate the numerical factors in final results for luminosity (\ref{eq:LumResFinAb}) and the perturbation $h^{GW}_{zz}$ (\ref{eq:rewrittenAgain}) of the space by the gravitatinal wave in $zz$ direction, which will be useful for real experiment. 

Now, we arrive to the expression for the luminosity as 
\begin{align}
\mathcal{L}_{\text{quad}}[\frac{erg}{s}]=&2.51\times 10^{-22}[\frac{{\rm s}^3}{{\rm kg}\, {\rm m}^2}]\frac{R_{t}[{\rm kg}/{\rm m}^3]}{\rho_{0}[{\rm g}/{\rm cm}^3]} P^2_{L}[\rm PW]\nonumber\\
&\times\left(-\frac{2}{3}+e^{-b_{I}}(b^2_{I}+1)\right)^2 \label{eq:NumericalLuminosityAblation}
\end{align} 
and we denote the part without the $b_{I}$ function as
\begin{align}
\mathcal{L}^{1}_{\text{quad}}[\frac{erg}{s}]=2.51\times 10^{-22}[\frac{{\rm s}^3}{{\rm kg}\, {\rm m}^2}]\frac{R_{t}[{\rm kg}/{\rm m}^3]}{\rho_{0}[{\rm g}/{\rm cm}^3]}P^2_{L}[\rm PW](-\frac{2}{3})^2. \label{eq:NumericalLuminosityAblationII}
\end{align} 

First, we will investigate the first time dependent part of (\ref{eq:DiagHpv}), we obtain
\begin{align}
h_{zz}=&2.817\times 10^{-39}[\frac{s^2}{{\rm kg}\, {\rm  m}}]\frac{1}{r[{\rm m}]}
\left(\frac{R_{t}[{\rm kg}/{\rm m}^3]}{\rho_{0}[{\rm g}/{\rm cm}^3]}\right)^{1/2}\nonumber\\
&\times E_{L}[\rm MJ](\frac{2}{3}+e^{-b_{I}}),\label{eq:NumericalHzzAbl}
\end{align}
and the second constant term is a new contribution to the result which depends on the geometry of the setup and the choice of $f_{1}$,
\begin{align}
h_{zz}^{sec}=&-6.201\times 10^{-43}[\frac{s^2}{\rm kg\, \text{m}}]\frac{S[{\rm cm}^2]R^{1/3}_{t}[\frac{{\rm kg}}{\text{m}}]^{1/3}}{r[\rm m]} I_{L}^{2/3}[\frac{\text{PW}}{\text{cm}^{2}}]^{2/3}\nonumber\\
&\times\left[f_{2}[\rm m](\frac{2}{3}+e^{-b_{I}})-\frac{3}{2}z_{L}[\rm m]e^{-b_{I}}\right]
\label{eq:NumericalHzzCOnst}.
\end{align}
The first expression in the second term has no physical meaning because we can make it zero by choosing different center of coordinate system with start at $d=f_{2}=0$. 
 
The value of $R_{t}$ for Carbon as a material for the target with $A=12,\,Z=6$ and wavelength $\lambda_{L}=0.35\times 10^{-4}$ cm, we will obtain $R_{t}=15.144[{\rm kg}/{\rm m}^3]$ from Eq.~\ref{eq:Pressure}.

For evaluation we will use the experimental values
\begin{align}
P_{L}=&\;0.5 \,\text{PW}, \,
\rho_{0}=\; 30 \,\text{mg}/\text{cm}^3, E_{L}=\;0.5 \,\text{MJ},
\tau =\;1 \,{\text{ns}},\label{eq:MomentEnEn2}
\end{align}
and the detection distance is $R=10$ m or equivalently $f_{2}=f=10$ m, $z_{L}=12$ m, parameters $a, b$ of the target foil are $a=b=1\,{\rm mm}=0.1\,{\rm cm}$ and therefore $I_{L}=50\,[\rm PW/{\rm cm}^2]$. The outgoing gravitational radiation has frequency $\nu_{g}=\; 1\, {\rm GHz}$ and wave length $\lambda_{g}=0.3$ m. The velocity $v_{r}=1.14\times 10^6 {\rm [m/s]}$, $b_{I}=0.2$ for time $t=10^{-9}$ s.

The final estimations for our expressions of the luminosity (\ref{eq:NumericalLuminosityAblation}) and the perturbation (\ref{eq:NumericalHzzAbl}) are:
\begin{align}
{\mathcal{L}}_{GW}&\simeq 3.61\times 10^{-20} [\text{erg}/\text{s}],\quad  h^{GW}_{zz}\simeq 4.7\times 10^{-39}.\label{eq:FinalResults}
\end{align}
The estimations are one lower lower in ${\mathcal{L}}_{GW}$ and three orders higher in $h^{GW}_{zz}$ compared to \cite{izestELINP2014, grossmannMeet2009l}. Our results contain new time dependent terms with function $b_{I}$ which modify the results and provide more precision.

The estimation for the constant term ${\mathcal{L}}^{1}_{GW}$ (\ref{eq:NumericalLuminosityAblationII}) and second term in $h^{sec}_{zz}$ (\ref{eq:NumericalHzzCOnst}) are
\begin{align}
{\mathcal{L}}^{1}_{GW}&=4.699\times 10^{-19}[\text{erg}/\text{s}],\quad
h_{zz}^{sec}=-2.45\times 10^{-39},\label{eq:L1}
\end{align}
which corresponds to the result  in  \cite{izestELINP2014, grossmannMeet2009l} but the order of ${\mathcal{L}}_{GW}$ is one order lower due to the $b_{I}$ terms. 

Interestingly, the second term (\ref{eq:NumericalHzzCOnst}) results in the estimation to a number $h_{zz}^{sec}=-2.45\times 10^{-39}$ which has the same order as (\ref{eq:FinalResults}). The term is partially of coordinate nature therefore we did not include it into final results.

We have derived and investigated generalized formulae for the luminosity (\ref{eq:LumResFinAb}) and the perturbation tensor $h_{zz}$ (\ref{eq:rewrittenAgain}) which newly shows non--trivial time dependence and depends on the function $b_{I}$ and on the laser wavelegth $\lambda_{g}$ through $R_{t}$.
 
\section{\label{sec:number3}The derivation of gravitational wave characteristics for piston model}

\subsection{The piston model}
\begin{figure}[h!]
\centering
\includegraphics[width=0.45\textwidth]{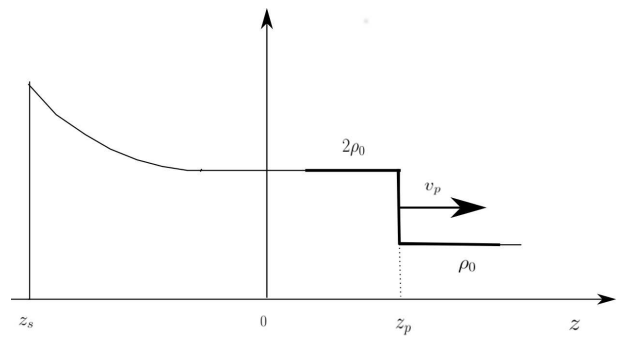}
\caption{\label{fig:wide4} The structure of the ion density profile of the piston caused by radiation pressure where the frame moves with the piston velocity $v_{p}$.}
\end{figure}

The recent progress in focal intensities of short-pulse lasers allows us to achieve intensities larger than $10^{20}$ W/$\text{cm}^2$ where the radiation pressure becomes the dominant effect in driving the motion of a particle in the material (target). The ponderomotive potential pushes the electrons steadily forward and the charge separation field forms a double layer (electrostatic shock or piston) propagating with $v_{p}$ where the ions are then accelerated forward. This strong electrostatic field forms a shocklike structure \cite{Naumova2009}. 

The use of circularly polarized laser light improves the efficiency of ponderomotive ion acceleration while avoiding the strong electron overheating. Then we will obtain quasi monoenergetic ion bunch in the homogeneous medium consisting of fast ions accelerated at the bottom of the channel with $20\%$ efficiency. The depth of penetration depends (in microns) on the laser fluence which should exceed tens of GJ$/\text{cm}^2$.

The model generates gravitational waves in THz frequency range with the duration of the pulse in picoseconds. The mass is accelerated with radiation pressure with circularly polarized pulse with intensity $I_{L}\geq 10^{21} \, \text{W}/\text{cm}^{2}$ which pushes the matter thanks to ponderomotive force. The matter is accelerated to the velocity $v_{p}$ which could be $10^9 \text{cm}/\text{s}$ and even more. 

\subsubsection{\label{eq:lim2}The limitations of the theory}
Let see whether the low velocity condition Eq.~(21) in \cite{Kadlecova2015} is satisfied for ablation model. The linear size of the source (focus size) is more than $d= 1{\mu}{\text m}= 10^{-6}\,{\text m}$  and the reduced generated wavelength is $\lambdabar=4.778\times 10^{-5} $~m for the gravitational wave length $\lambda_{g}=300$ $\mu$m.
The comparison Eq.~(22) in \cite{Kadlecova2015} results into
\begin{equation}
0.021 \ll 1.\label{eq:compLambdaP}
\end{equation}
The low velocity condition is still satisfied for the piston model experiment, while
we have a limit for the size of the target for the piston model.

\subsubsection{Set up of the experiment}
The set up for the experiment is visible in Fig.~\ref{fig:wide4}. The target is positioned at the start of the coordinate system $x,\,y,\,z$ and we expect that the depth of hole boring is very small. The detector is positioned in the same distance as in the previous models, in the distance $z_{D}=10$ m.

The material is accelerated in the direction of the $z$ coordinate. The function of the shock position is again taken 
\begin{align}
z_{p}(t)= v_{p}t + d, \label{eq:zp}
\end{align}
like in the previous models, see \cite{Kadlecova2015} and (\ref{eq:zr}) for comparison.

The velocity of a piston is denoted as
\begin{align}
v_{p}\simeq \sqrt{\frac{I_{L}}{c\rho_{0}}}, \label{eq:vp}
\end{align}
where $\rho_{0}$ is material density and $I_{L}$ is the intensity of the laser in PW/$\text{cm}^2$.
We  have denoted the velocity as (\ref{eq:vp}) and we assume that for $t=0$, $z_{s}(0)=0$, therefore $d=0$ according the the Fig. \ref{fig:wide4}.

The time when the radiation reaches the detector is defined as
\begin{align}
t_{D}= \frac{z_{D}}{v_{P}}. \label{eq:tD}
\end{align}

Again, we will calculate everything with general function $z_{p}(t)$ and then we will substitute the explicit function (\ref{eq:zp}) at convenient places which might be useful for other forms of $z_{p}(t)$.
 
The basic input for the calculation is the density profile from Fig.~\ref{fig:wide4}. The step function for the density profile can be written as
\begin{equation}
\rho(t,\bf{x})=
\begin{cases}
\;2\rho_{0} &\quad if \quad z < z_{p},\label{eq:densityShock}\\
\;\rho_{0} &\quad if \quad z > z_{p}.
\end{cases}
\end{equation}
The first step in the calculation is the mass moment derivation.

\subsubsection{The mass moment}
The values for integration of the density (\ref{eq:densityShock}) in Eq.~(11) in \cite{Kadlecova2015} are $x\in <0,a)$, $y\in <0,b)$ and $z\in<0,z_{D}>$ which splits into $<0,z_{p})$ and $(z_{s},z_{D}>$. The mass moment diagonal components then read
\begin{align}
M_{xx}&=\frac{Sa^2}{3}\rho_{0}\left(z_{p}+z_{D}\right),\quad
M_{yy}=\frac{Sb^2}{3}\rho_{0}\left(z_{p}+z_{D}\right),\nonumber\\
M_{zz}&=\frac{S}{3}\rho_{0}\left(z_{p}^3+z^3_{D}\right),\label{eq:componentsMassMomentDiag}
\end{align}
and non--diagonal components $M_{xy}, M_{yz}, M_{xz}$,
\begin{align}
M_{xy}&=\frac{S^2}{4}\rho_{0}\left(z_{p}+z_{D}\right),\quad M_{yz}=\frac{Sb}{4}\rho_{0}\left(z^2_{p}+z^2_{D}\right),\nonumber\\
M_{xz}&=\frac{Sa}{4}\rho_{0}\left(z^2_{p}+z^2_{D}\right).\label{eq:componentsMassMomentOff}
\end{align}
These semi--results will be usefull for the polarization because it shows that it is sometimes more convenient to use the mass moment  for calculations instead of the quadrupole moment. 

\subsubsection{The quadrupole moment}
The non--diagonal components $I_{xy}, I_{yz}, I_{xz}$ are
\begin{align}
I_{xy}&=M_{xy},\quad I_{yz}=M_{yz},\quad I_{xz}=M_{xz}.\label{eq:componentsQuadrMomentOff}
\end{align}
The diagonal components $I_{ii}=M_{ii}-\frac{1}{3}Tr M$ read
\begin{align}
I_{xx}&=\frac{S\rho_{0}}{9}\left\lbrace -z_{p}^3+(2a^2-b^2)(z_{p}+z_{D})- z^3_{D}\right\rbrace,\nonumber\\
I_{yy}&=\frac{S\rho_{0}}{9}\left\lbrace -z_{p}^3+(2b^2-a^2)(z_{p}+z_{D})- z^3_{D}\right\rbrace,\nonumber\\
I_{zz}&=\frac{S\rho_{0}}{9}\left\{2z^3_{p}-(a^2+b^2)(z_{p}+z_{D})+z^3_{D}\right\}.\label{eq:componentsQuadMomentDiag}
\end{align}
The functional dependence is almost the same as in the previous models thanks to the
linearity of the function $z_{p}(t)$. The component $I_{zz}$ then becomes explicitly
\begin{align}
I_{zz}&=\frac{S\rho_{0}}{9}\left\{2v^3_{p}t^3-(a^2+b^2)v_{p}t+z_{D}(2z^2_{D}-(a^2+b^2))\right\}.\label{eq:IzzWithSubst}
\end{align}
The quadrupole moment in the $zz$ direction is given by a cubic polynomial in $t$ time variable. 

When we compare our result (\ref{eq:IzzWithSubst}) with \cite{izestELINP2014, grossmannMeet2009l} we observe (again) that just the most dominant term was used for their calculations. The other terms are new, linear and constant terms. The geometry of the setup influences the quadrupole moment from the linear term and lower.
The derivatives of the quadrupole moment and mass moment are listed in Appendix A, the derivatives with dependence on $z_{p}$ in (\ref{A2}) and with substitution of $z_{p}$ in (\ref{A3}).

\subsubsection{The analytical form of perturbation and luminosity}
Now, we calculate the components of the perturbation tensor according to Eq.~(9) in \cite{Kadlecova2015} without projector $\Lambda_{ij,kl}({\bf n})$. In other words, we got the components of the perturbation tenzor in general form, the components read
\begin{align}
h_{xx}&=\frac{2G}{9rc^4}S\rho_{0}\left\{(2a^2-b^2)\ddot{z}_{p}-(z^3_{p})^{\ddot{}}\right\},\nonumber\\
h_{yy}&=\frac{2G}{9rc^4}S\rho_{0}\left\{(2b^2-a^2)\ddot{z}_{p}-(z^3_{p})^{\ddot{}}\right\},\label{eq:DiagHpert}\\
h_{zz}&=\frac{2G}{9rc^4}S\rho_{0}\left\{2(z^3_{p})^{\ddot{}}-(a^2+b^2)\ddot{z}_{p}\right\},\nonumber
\end{align}
and the non-diagonal terms are
\begin{align}
h_{xy}&=\frac{G}{2rc^4}S^2\rho_{0}\ddot{z}_{p},\quad h_{xz}=\frac{G}{2rc^4}Sa\rho_{0}(z^2_{p})^{\ddot{}},\nonumber\\ 
h_{yz}&=\frac{G}{2rc^4}Sb\rho_{0}(z^2_{p})^{\ddot{}}.\label{eq:NodiagHpert}
\end{align}

The perturbation tensor with substitution of $z_{p}(t)$ reads
\begin{align}
h_{xx}&=-\frac{4G}{3rc^4}S\rho_{0}v^3_{p}t,
h_{yy}=-\frac{4G}{3rc^4}S\rho_{0}v^3_{p}t,\nonumber\\
 h_{zz}&=\frac{8G}{3rc^4}S\rho_{0}v^3_{p}t,\label{eq:Dvp}
\end{align}
and the non-diagonal terms are
\begin{align}
h_{xy}=0,\quad h_{xz}=\frac{G}{rc^4}Sa\rho_{0}v^2_{p},\quad h_{yz}=\frac{G}{rc^4}Sb\rho_{0}v^2_{p},\label{eq:Nodiagvp}
\end{align}
where we used the derivatives of $z_{p}$ listed in Appendix \ref{App:A1}.

After substituting the quadrupole moment components into Eq.~(10) in \cite{Kadlecova2015}, we get general expression as
\begin{align}
\mathcal{L}_{\text{quad}}&=\frac{G}{405c^5}S^2\rho^2_{0}\left\{6[(z^3_{p})^{\dddot{}}]^2-6\dddot{z}_{p}(z^3_{s})^{\dddot{}}(a^2+b^2)\right.\nonumber\\
 +&\left.(\dddot{z}_{p})^2[(a^2+b^2)^2+(2a^2-b^2)^2+(2b^2-a^2)^2+\frac{81}{16}S^2]\right.\nonumber\\
 +&\left.\frac{81}{16}(a^2+b^2)[(z^2_{p})^{\dddot{}}]^2\right\}. \label{eq:LumRes2p}
\end{align}
The explicit substitution $z_{p}$ simplifies the expression Eq.~(10) in \cite{Kadlecova2015} that just the diagonal components of quadrupole moment contribute to the result, see (\ref{eq:firstDerQuadrDiag}). The expression (\ref{eq:LumRes2p})
further simplifies to 
\begin{align}
\mathcal{L}_{\text{quad}}=&\frac{8G}{15c^5}S^2\rho_{0}^2v^6_{p}. \label{eq:LumResvp}
\end{align}

After inserting (\ref{eq:vp}) and (\ref{eq:Pressure}), we will obtain the final expression for luminosity of gravitational radiation,
\begin{align}
\mathcal{L}_{\text{quad}}=\frac{8}{15}\frac{G}{c^5}\frac{1}{S\rho_{0}}\left(\frac{P_{L}}{c}\right)^3, \label{eq:LumResFin}
\end{align}
where we have used the pressure (\ref{eq:Pressure}).

The luminosity then depends on the power of the laser, the density of the material and the laser wavelength and the surface of the focal spot $S$. The numerical factor in front of the fraction for estimation will be presented in the next subsection.

The perturbation component $h^{GW}_{zz}$ becomes using 
(\ref{eq:vp}), (\ref{eq:Pressure}) and (\ref{eq:pressureAndenergy}),
\begin{align}
h_{zz}&=\frac{8G}{rc^4}\frac{1}{\sqrt{S\rho_{0}}}\left(\frac{P_{L}}{c}\right)^{3/2}t.\label{eq:DiagHpv}
\end{align}

This is the final formula for the perturbation of the space by gravitational wave in the $zz$ direction. The formula has different power of laser power than the previous models. The value of the perturbation decreases with the distance as $1/r$ and will be zero in the infinity. The numerical factors will be evaluated in the next subsection for specific values for an experiment.

\subsubsection{\label{subsec:estimate}The estimations for the $h_{\mu\nu}$ and $\mathcal{L}_{\text{quad}}$ for real experiment}
We will evaluate the numerical factors in final results for luminosity (\ref{eq:LumResFin}) and the perturbation $h^{GW}_{zz}$ of the space by the gravitatinal wave in $zz$ direction, (\ref{eq:DiagHpv}), which will be useful for real experiment. 
Now, we arrive to the expression for the luminosity as 
\begin{align}
\mathcal{L}_{\text{quad}}[\frac{erg}{s}]=5.572\times 10^{-30}[\frac{{\rm s}^6}{{\rm kg}\, {\rm m}^5}]\frac{P^3_{L}[{\rm PW}]}{S[{\rm m}^2]\rho_{0}[{\rm g}/{\rm cm}^3]}. \label{eq:NumericalLuminosity}
\end{align} 

Similarly to the previous case, we obtain
\begin{align}
h_{zz}=&2.2267\times 10^{-35}[\frac{kg^{7/2}}{{\rm s^2}\, {\rm m^{5/2}}}]\frac{1}{r[{\rm m}]}
\frac{P^{3/2}_{L}[PW]t[{\rm ps}]}{\left(S[{\rm m^2}]\rho_{0}[{\rm g}/{\rm cm}^3]\right)^{1/2}}.\label{eq:NumericalHzz}
\end{align}

When we substitute achievable laser parameters into expressions for luminosity and the perturbation we will get the estimations for the experiment:
\begin{align}
P_{L}=&\;7\,\text{PW},\,\rho_{0}=\, 1\,\text{g}/\text{cm}^3,\,\Phi=\;30 \,\mu\text{m},\;\tau =\;1 \,{\text{ps}},\label{eq:MomentEnEn}
\end{align}
and the detection distance is again $R=10$ m and $S=\Phi^2\pi/4$ where $\Phi$ is diameter of the target. The detection distance is $R=10$ m or equivalently $f_{2}=f=10$ m, $z_{L}=12$ m, parameters $a, b$ of the target foil are $a=b=1\,{\rm \mu m}=1\times 10^{-6}\,{\rm m}$ and therefore $I_{L}=7\times 10^{8}\,[\rm PW/{\rm cm}^2]$ and the velocity $v_{r}=153008\,{\rm [km/s]}$.

The wavelenght of the gravitational wave is $\lambda_{g}=300\,{\rm \mu m}$ and the frequency is $\nu_{g}=1$ THz. 

The final estimations for the luminosity and the perturbations are:
\begin{equation}
{\mathcal{L}}_{GW}\simeq \,2.704\times 10^{-18} [\text{erg}/\text{s}], \, h^{GW}_{zz}\simeq 3\times 10^{-43}.\label{eq:PistonFinPh}
\end{equation}
The estimates for $\mathcal{L}_{GW}$ and $h_{\mu\nu}$ are one order lower than the result in \cite{izestELINP2014, grossmannMeet2009l}.  

\section{\label{sec:number4}The polarization of gravitational waves}
In this section, we are going to investigate the two polarization modes of the gravitational waves which are generated by ablation and piston models. We derive the amplitudes of the gravitational wave in two independent modes, $+$ and $-$, and focus on their interpretation which would be useful for real experiment conditions while we will refer to the theory part in the first part of this paper \cite{Kadlecova2015}.

\subsection{The $x$, $y$ and $z$ directions of the wave vector for ablation model}
First, we are going to investigate the gravitational perturbations in the direction of the propagation, in the $z$--coordinate.

\subsubsection{\label{sub:z}The wave propagation in the $z$--direction}
The $h^{TT}_{ij}$ Eq.~(9) in \cite{Kadlecova2015} has then the only non--vanishing components
\begin{align}
h^{TT}_{xx}&=-h^{TT}_{yy}=\text{Re}\left\{A_{+}e^{-i\omega(t+z/c)}\right\},\nonumber\\
h^{TT}_{xy}&=h^{TT}_{yx}=\text{Re}\left\{A_{\times}e^{-i\omega(t+z/c)}\right\},\label{eq:hTTcomponentsAbl}
\end{align}
for the wave propagation vector in the $z$--direction $n=(0,0,-1)$.

The waves are linearly polarized in the direction of propagation as in the case of shock wave model \cite{Kadlecova2015}. We obtain the amplitudes of the polarization modes for the ablation model in the form, Eq.~(49) in \cite{Kadlecova2015} 
then we use the mass moments expressed in terms of derivatives of function $z$,
\begin{align}
A^{a}_{+}&=\frac{4}{3r}\frac{G}{c^4}S\rho_{0}(b^2-a^2)z^2_{L}\frac{e^{-b_{I}}}{z^3_{r}}=0,\nonumber\\
A^{a}_{\times}&=-\frac{2}{r}\frac{G}{c^4}S^2\rho_{0}z^2_{L}\frac{e^{-b_{I}}}{z^3_{r}}\label{eq:hkz1a}.
\end{align} 

\begin{figure}[h]
\centering
\includegraphics[width=0.45\textwidth]{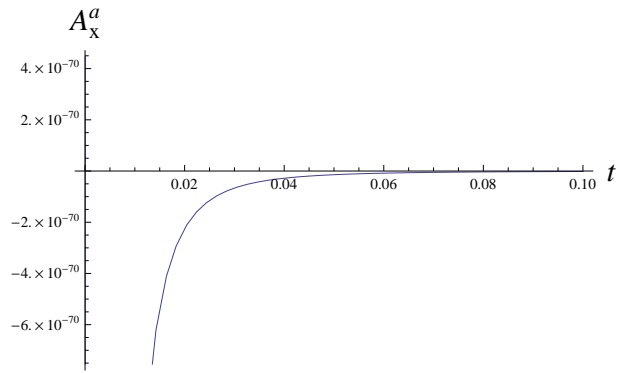}
\caption{\label{fig:AmplitudeZ} The amplitudes $A^{a}_{\times}$ (\ref{eq:hkz1a}) is pictured with dependence on time $t[s]$. The amplitude approaches zero quickly.}
\end{figure}

The time dependency is hidden in $z_{r}$ (\ref{eq:zr}). Contrary to the shock wave model \cite{Kadlecova2015} and piston model (\ref{sub:Piston}) the amplitudes do not vanish but are quite small O($\frac{e^{-b_{I}}}{z_{r}^3}$) and vanish as $t\rightarrow \infty$ or $r\rightarrow \infty$.  The amplitude $A^{a}_{+}=0$ because of our choice of square target $b^2-a^2=0$. The remaining amplitude $A^{a}_{\times}$ is pictured in Fig.~\ref{fig:AmplitudeZ}, where we observe that the amplitude approaches zero quickly. Therefore waves do radiate along the $z$ axis in which the motion occurs but very weakly. It is surprising result because in the linear gravitation such waves do not exist, just the transversal ones. It is the consequence of the non--conservation of mass by the ablation model and the finite integration boundary $z_{L}$.

The gravitational radiation is strongly non--zero in the other directions, for example in the direction of the $x$ and $y$ axes, see the next subsections.  

\subsubsection{\label{sub:x}The wave propagation in the $x$--direction}
The $h^{TT}_{ij}$ Eq.~(9) in \cite{Kadlecova2015} has the only non--vanishing components for the wave vector in the $x$--direction $n=(1,0,0)$,
\begin{align}
h^{TT}_{yy}&=-h^{TT}_{zz}=\text{Re}\left\{A_{+}e^{-i\omega(t-x/c)}\right\},\nonumber\\
h^{TT}_{zy}&=h^{TT}_{yz}=\text{Re}\left\{A_{\times}e^{-i\omega(t-x/c)}\right\}.\label{eq:hTTcomponentsX}
\end{align}

The waves are linearly polarized as in the previous case. We obtain the amplitudes of the polarization modes, Eq.~(54) in \cite{Kadlecova2015} 
then we use the mass moments expressed in terms of derivatives of function $z$, the amplitudes read as follows,
\begin{align}
A^{a}_{+}&=\frac{4}{r}\frac{G}{c^4}S\rho_{0}\left[-6z_{r}(\dot{z}_{r})^2(-\frac{2}{3}+e^{-b_{I}}(b^2_{I}+1))\right.\nonumber\\
+&4\left.\dot{z}_{r}e^{-b_{I}}z_{L}(\frac{z_{L}}{z_{r}}-1)+12(\dot{z}_{r})^2z_{L}e^{-b_{I}}\right.\nonumber\\
-&\left.\frac{e^{-b_{I}}}{z_{r}}z^2_{L}(1-\frac{2z_{L}}{z_{r}})-\frac{e^{-b_{I}}}{3z^3_{r}}z^2_{L}b^2\right], \label{eq:hpx1a}\\
A^{a}_{\times}&=\frac{4}{r}\frac{G}{c^4}Sb\rho_{0}\left[(z^2_{r})^{\ddot{}}(\frac{1}{2}+b_{I}e^{-b_{I}})+\frac{z^2_{L}}{z^3_{r}}e^{-b_{I}}(2(z^2_{r}-(z_{r}+z_{L}))\right].\label{eq:hkx1a}
\end{align}

We have obtained non--zero amplitudes for both '$+$' and '$\times$' polarization modes. The amplitudes depend on the focus area $S$, the density of the material $\rho_{0}$, the velocity of the ions $v_{r}$ and constant $z_{L}$. The amplitudes vanish as the radial distance $r\rightarrow\infty$ and they decrease like $1/r$.

Importantly, both amplitudes of '$+$' and '$\times$' polarization are time dependent. The dependency originates from the expression $b_{I}$ (\ref{eq:mztf}) which was not present in the shock wave model and in fact generalizes the results of the shock wave model \cite{Kadlecova2015}. The amplitude for '$\times$' polarization was not time dependent.

We observe that the terms containing $b_{I}$ in the numerator contribute less in the limit $t \rightarrow \infty$, such has $\lim_{t\to\infty}e^{-b_{I}}=e^{-1}$ and $\lim_{t\to\infty}b_{I}=1$, the terms as $\frac{e^{-b_{I}}}{z^{k}_{r}}$, where $k=1, 2, 3$, vanish in the limit. The most dominant terms remain the first terms in the expressions for the amplitudes (\ref{eq:hpx1a}) and (\ref{eq:hkx1a}) which have are 
functionaly similar character, except the terms with $b_{I}$, as the shock wave model.

When the radiation reaches the detector at $t_{det}=f_{2}/v_{r}$, the most dominant term in $A^{a}_{+}$ vanishes, the last two diverge since the division by $0$. The $A^{a}_{\times}$ has just the first term non--divergent.

\begin{figure}[h]
\centering
\includegraphics[width=0.45\textwidth]{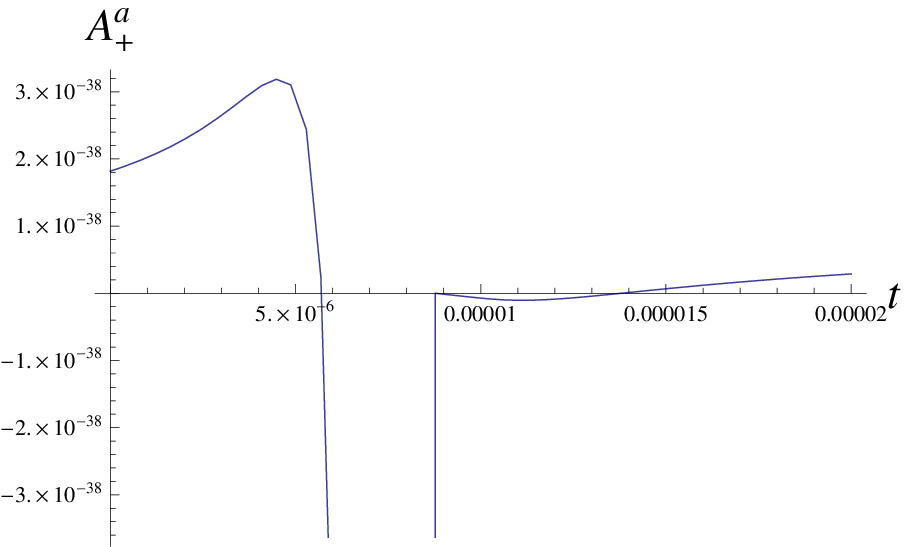}
\includegraphics[width=0.45\textwidth]{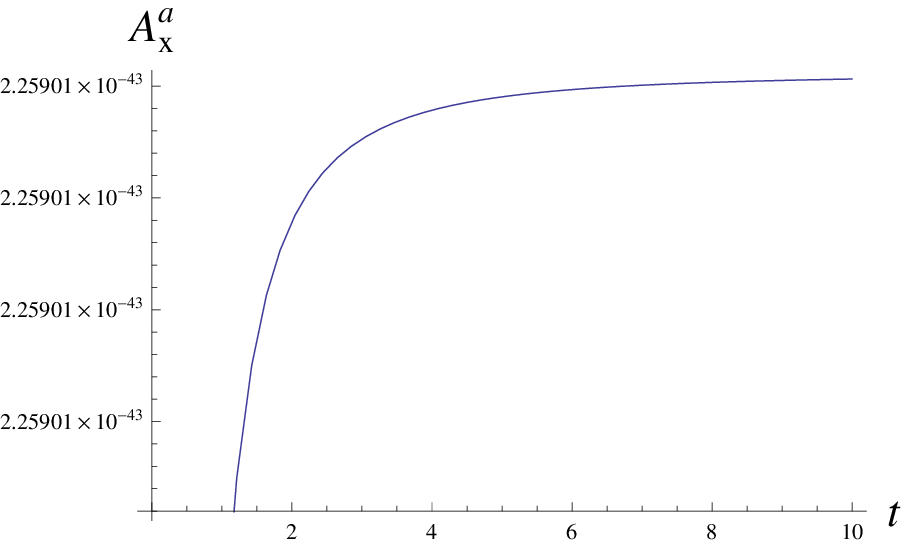}
\caption{\label{fig:AmplitudeX} The amplitudes $A^{a}_{+}$ (\ref{eq:hpx1aa}) and $A^{a}_{\times}$ (\ref{eq:hkx1aa}) are pictured in dependence on time $t[s]$. The amplitudes do not vanish in time due to fact that mass is not conserved by the ablation model.}
\end{figure}

The amplitudes then reduce to 
\begin{align}
A^{a}_{+}&=-\frac{8}{r}\frac{G}{c^4}S\rho_{0}v^2_{r}\left[3z_{r}(-\frac{2}{3}+e^{-b_{I}}(b^2_{I}+1))-2e^{-b_{I}}z_{L}(3 -\frac{1}{\dot{z}_{r}})\right], \label{eq:hpx1aa}\\
A^{a}_{\times}&=\frac{4}{r}\frac{G}{c^4}Sb\rho_{0}\left[2v^2_{r}(\frac{1}{2}+b_{I}e^{-b_{I}})\label{eq:hkx1aa}\right],
\end{align}
while we have omitted the terms of type $e^{-b_{I}}/z_{r}$ which diverge for our choice of the start of coordinate system and have smaller additional contribution than the remaining terms. The amplitudes are depicted in the Fig.~\ref{fig:AmplitudeX} for experimental values specified in estimations part (\ref{subsec:estimate}). The amplitude $A^{a}_{+}$ shows jump down at $t_{det}$ because of the $zr=0$ and then grows like the amplitude $A^{a}_{\times}$.  The amplitude $A^{a}_{+}$ shows open profile function which continues to $\infty$. Correctly, the function should close down because GW loses its energy. The opened function is again caused by the mass non--conservation in the ablation model.
We will investigate the influence of the wave on test particles in Section (\ref{sec:number6}).

\subsubsection{\label{sub:y}The wave propagation in the $y$--direction}
The last direction we are going to investigate is the $y$-direction transversal to the direction of motion in $z$--coordinate.
The perturbation tenzor Eq.~(9) in \cite{Kadlecova2015} for the wave vector in the $y$--direction $n=(0,1,0)$ reads
\begin{align}
h^{TT}_{xx}&=-h^{TT}_{zz}=\text{Re}\left\{A_{+}e^{-i\omega(t-y/c)}\right\},\nonumber\\
h^{TT}_{zx}&=h^{TT}_{xz}=\text{Re}\left\{A_{\times}e^{-i\omega(t-y/c)}\right\}.\label{eq:hTTcomponentsY}
\end{align}

Again, the waves are linearly polarized as in the previous cases. The amplitudes of the polarization modes become, Eq.~(59) in \cite{Kadlecova2015}
then we use the mass moments expressed in terms of derivatives of function $z$,
\begin{align}
A^{a}_{+}&=\frac{4}{r}\frac{G}{c^4}S\rho_{0}\left[-6z_{r}(\dot{z}_{r})^2(-\frac{2}{3}+e^{-b_{I}}(b^2_{I}+1))\right.\nonumber\\
&+4\left.\dot{z}_{r}e^{-b_{I}}z_{L}(\frac{z_{L}}{z_{r}}-1)+12(\dot{z}_{r})^2z_{L}e^{-b_{I}}\right.\nonumber\\
&\left.-\frac{e^{-b_{I}}}{z_{r}}z^2_{L}(1-\frac{2z_{L}}{z_{r}})-\frac{e^{-b_{I}}}{3z^3_{r}}z^2_{L}a^2\right], \label{eq:hpy1a}\\
A^{a}_{\times}&=-\frac{4}{r}\frac{G}{c^4}Sa\rho_{0}\left[(z^2_{r})^{\ddot{}}(\frac{1}{2}+b_{I}e^{-b_{I}})\right.\nonumber\\
&+\left.\frac{z^2_{L}}{z^3_{r}}e^{-b_{I}}(2(z^2_{r})^{\dot{}}-(z_{r}+z_{L}))\right]\label{eq:hky1a}.
\end{align}

\begin{figure}[h]
\centering
\includegraphics[width=0.5\textwidth]{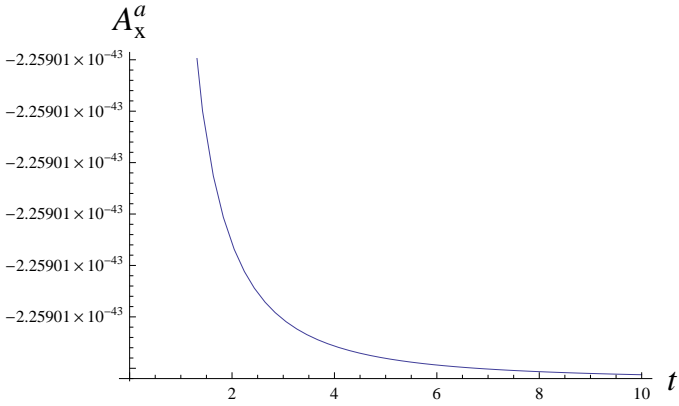}
\caption{\label{fig:AmplitudeY} The amplitude $A^{a}_{\times}$ (\ref{eq:hky1aa}) is pictured in dependence on time $t[s]$. The image for $A^{a}_{+}$ in Fig.~\ref{fig:AmplitudeX} is the same for this case.}
\end{figure}

The resulting amplitudes $A^{a}_{+}$ and $A^{a}_{-}$ have the form like in the direction $x$ (\ref{eq:hpx1a}) and (\ref{eq:hkx1a}) apart from the sign in $A^{a}_{\times}$ and parameter $a$ instead $b$. Importantly, the $A^{a}_{+}$ and $A^{a}_{\times}$  amplitudes are dependent on time. The results have the same character as in the previous case.  The amplitudes vanish as the radial distance $r\rightarrow\infty$ and decrease as $1/r$.

\begin{align}
A^{a}_{+}&=-\frac{8}{r}\frac{G}{c^4}S\rho_{0}v^2_{r}\left[3z_{r}(-\frac{2}{3}+e^{-b_{I}}(b^2_{I}+1))\right.\nonumber\\
&\left.-2e^{-b_{I}}z_{L}(3 -\frac{1}{\dot{z}_{r}})\right], \label{eq:hpy1aa}\\
A^{a}_{\times}&=-\frac{4}{r}\frac{G}{c^4}Sa\rho_{0}\left[2v^2_{r}(\frac{1}{2}+b_{I}e^{-b_{I}})\right]\label{eq:hky1aa},
\end{align}
while we have omitted the terms of type $e^{-b_{I}}/z_{r}$ which diverge for our choice of the start of coordinate system and have smaller additional contribution than the remaining terms. The amplitudes are depicted in Fig.~\ref{fig:AmplitudeY} which is just rotated Fig.~\ref{fig:AmplitudeX} because of the minus sign in (\ref{eq:hky1aa}).

The amplitudes of radiation  and the radiative characterictics of the radiation are one of the main results of this paper. 

\subsection{\label{sub:Piston}The $x$, $y$ and $z$ directions of the wave vector for piston model}
First, we are going to investigate the gravitational perturbations in the direction of the propagation, in the $z$--coordinate.
 
\subsubsection{\label{sub:zz}The wave propagation in the $z$--direction}
The $h^{TT}_{ij}$ Eq.~(9) in \cite{Kadlecova2015} has then the only non--vanishing components (\ref{eq:hTTcomponentsAbl}) for the wave propagation vector in the $z$--direction $n=(0,0,1)$.
The amplitudes are given by Eq.~(49) in \cite{Kadlecova2015},
after substituting the $z_{p}$ (\ref{eq:zp}) read as follows,
\begin{align}
A^{p}_{+}&=\frac{1}{r}\frac{G}{3c^4}S\rho_{0}\ddot{z}_{p}(a^2-b^2)=0, \quad
A^{p}_{\times}=-\frac{1}{2r}\frac{G}{c^4}S^2\rho_{0}\ddot{z}_{p}=0 \label{eq:hkz1}.
\end{align} 

Therefore the radiation ${h^{TT}_{ij}}$ is vanishing for the orientation of the wave vector into the direction of motion of the experiment. The waves do not radiate along the $z$ axis.

\subsubsection{\label{sub:xx}The wave propagation in the $x$--direction}
The $h^{TT}_{ij}$ Eq.~(9) in \cite{Kadlecova2015} has the only non--vanishing components for the wave vector in the $x$--direction $n=(1,0,0)$ (\ref{eq:hTTcomponentsX})
where the amplitude are given by Eq.~(54) in \cite{Kadlecova2015} and after substitution to $z_{p}$ we get,
\begin{align}
A^{p}_{+}&=\frac{1}{3r}\frac{G}{c^4}S\rho_{0}\left[b^2\ddot{z}_{p}-({z_{p}^{3}})^{\ddot{}}\right]=-\frac{2}{r}\frac{G}{c^4}S\rho_{0}v_{p}^3t\,, \label{eq:hpx1p}\\
A^{p}_{\times}&=\frac{1}{2r}\frac{G}{c^4}S\rho_{0}b(z^{2}_{p})^{\ddot{}}=\frac{1}{r}\frac{G}{c^4}S\rho_{0}bv_{p}^2\label{eq:hkxp}.
\end{align}

\subsubsection{\label{sub:yy}The wave propagation in the $y$--direction}
The perturbation tenzor in $TT$ calibration Eq.~(9) in \cite{Kadlecova2015}
for the wave vector in the $y$--direction $n=(0,1,0)$ are (\ref{eq:hTTcomponentsY}).
The amplitudes are given by Eq.~(59) in \cite{Kadlecova2015} and after substitution for $z_{r}$ we get,
\begin{align}
A^{p}_{+}&=\frac{1}{3r}\frac{G}{c^4}S\rho_{0}\left[a^2\ddot{z}_{p}-({z_{p}}^{3})^{\ddot{}}\,\right]=-\frac{2}{r}\frac{G}{c^4}S\rho_{0}v_{p}^3 t, \label{eq:hpy1}\\
A^{p}_{\times}&=-\frac{1}{2r}\frac{G}{c^4}S\rho_{0}a({z_{s}}^{2})=-\frac{1}{r}\frac{G}{c^4}S\rho_{0}av_{p}^2\label{eq:hky1}.
\end{align}

The resulting amplitudes $A^{p}_{+}$ and $A^{p}_{-}$ have the form as in the direction $x$ (\ref{eq:hpx1p}) and (\ref{eq:hkxp}) apart from the sign in $A^{p}_{\times}$. Importantly, the $A^{p}_{+}$ amplitude depends linearly on time and again the other one $A^{p}_{\times}$ is constant in time. The results have the same character as in the previous case and correspond to results for shock wave model \cite{Kadlecova2015}.  The amplitudes vanish as the radial distance $r\rightarrow\infty$ and decrease as $1/r$.

The GW amplitudes are the main result of the paper. 

\subsection{The general direction of the wave vector}
Finally, we are going to investigate the amplitudes with the general wave vector of propagation. The general direction of the wave propagation can be expressed in the spherical coordinates as $n = (\sin\theta\sin\phi,\,\sin\theta\cos\phi,\,\cos\theta),$
and the perturbation tenzor can be obtained via Eq.~(9) in \cite{Kadlecova2015} and the projector $\Lambda_{ij,kl}$.

\subsubsection{The case of ablation model}
The general expressions for the two modes of polarizations are Eq.~(62-63) in \cite{Kadlecova2015}, (\cite{MaggioreBook}),
Afterwards we use the mass moments expressed in terms of derivatives of function $z$, the amplitudes  read as follows,
\begin{align}
&A^{a}_{+}(t;\theta,\phi)=\frac{1}{r}\frac{G}{c^4}S\rho_{0}\nonumber\\
&\times\left[-\frac{4}{3}\frac{e^{-b_{I}}}{z^3_{r}}z^2_{L}\left[a^2(\cos^2\phi-\sin^2\phi\cos^2\theta)\right.+\left.b^2(\sin^2\phi-\cos^2\phi\cos^2\theta)\right]\right.\nonumber\\
&\left.+2\sin{2\theta}(a\sin\phi+b\cos\phi)[(z^2_{r})^{\ddot{}}(\frac{1}{2}+b_{I}e^{-b_{I}})\right.\nonumber\\
&+\left.\frac{z^2_{L}}{z^3_{r}}e^{-b_{I}}(2(z^2_{r})^{\dot{}}-(z_{r}+z_{L})]\right.\label{eq:hknAbl1}\\
&\left.-Sz^2_{L}\frac{e^{-b_{I}}}{z^3_{r}}\sin 2\phi(1+\cos^2\theta)\right.\nonumber\\
&-\left.4\sin^2\theta\left[6z_{r}(\dot{z}_{r})^2(-\frac{2}{3}+e^{-b_{I}}(b^2_{I}+1))\right.\right.\nonumber\\
&\left.\left.-12(\dot{z}_{r})^{2}e^{-b_{I}}z_{L}-4e^{-b_{I}}z_{L}\dot{z}_{r}(\frac{z_{L}}{z_{r}}-1)\right.+\left.\frac{e^{-b_{I}}}{z_{r}}z^2_{L}(1-\frac{2z_{L}}{z_{r}})\right]\right],\nonumber\\
&A^{a}_{\times}(t;\theta,\phi)=\frac{2}{r}\frac{G}{c^4}S\rho_{0}\left[-2\sin\theta(a\cos\phi-b\sin\phi)\right.\nonumber\\
&\times\left.\left((z_{r}^2)^{\ddot{}}(\frac{1}{2}+b_{I}e^{-b_{I}})+\frac{z^2_{L}}{z^3_{r}}e^{-b_{I}}(2(z_{r}^2)^{\dot{}}-(z_{r}+z_{L}))\right)\right.\nonumber\\
&-\left.\frac{e^{-b_{I}}}{z^3_{r}}z^2_{L}\cos\theta\left(\frac{2}{3}(a^2-b^2)\sin{2\phi}+\frac{1}{2}S\cos{2\phi}\right)\right].\label{eq:hknAbl2}
\end{align}

We obtained the amplitudes of two independent polarization modes with the general wave vector of propagation. The character of the amplitudes resembles the results from two previous cases, the amplitude $A^{a}_{+}$ is linearly time dependent and the $A^{a}_{\times}$ is constant in time. The amplitudes vanish as the radial distance $r\rightarrow\infty$ and decreases as $1/r$.

We will obtain the three previous cases as subcases of these general amplitudes. The case ${\bf n}=z$ (\ref{sub:z}) for $\theta=0^{\circ},\,\phi=0^{\circ}$, the case ${\bf n}=x$ (\ref{sub:x}) can be obtained for $\theta=90^{\circ},\, \phi=90^{\circ}$ and case ${\bf n}=y$ (\ref{sub:y}) for $\theta=90^{\circ},\,\phi=0^{\circ}$.

To visualize the amplitudes we will omit the terms of type $e^{-b_{I}}/z_{r}$, 
\begin{align}
A^{a}_{+}(t;\theta,\phi)&=\frac{4}{r}\frac{G}{c^4}S\rho_{0}v^2_{r}\left[(\frac{1}{2}+b_{I}e^{-b_{I}})\sin{2\theta}(a\sin\phi+b\cos\phi)\right.\nonumber\\
&\left.-3\sin^2\theta\left[3(-v_{r}t+f_{2})(-\frac{2}{3}+e^{-b_{I}}(b^2_{I}+1)\right.\right.\nonumber\\
&\left.\left.-2z_{L}e^{-b_{I}}(3-\frac{1}{\dot{z}_{r}})\right]\right],\label{eq:hknAbl2b}\\
A^{a}_{\times}(t;\theta,\phi)&=-\frac{4}{r}\frac{G}{c^4}S\rho_{0}v^2_{r}\left[(\frac{1}{2}+b_{I}e^{-b_{I}})\sin\theta(a\cos\phi-b\sin\phi)\right].\label{eq:hknAbl2a}
\end{align}

\begin{figure}[h]
\centering
\includegraphics[width=0.5\textwidth]{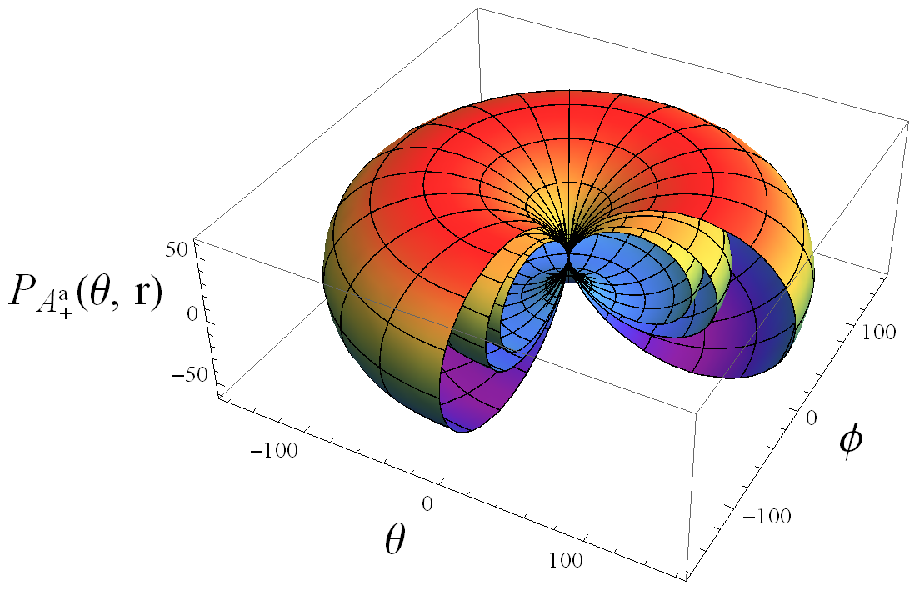}
\includegraphics[width=0.4\textwidth]{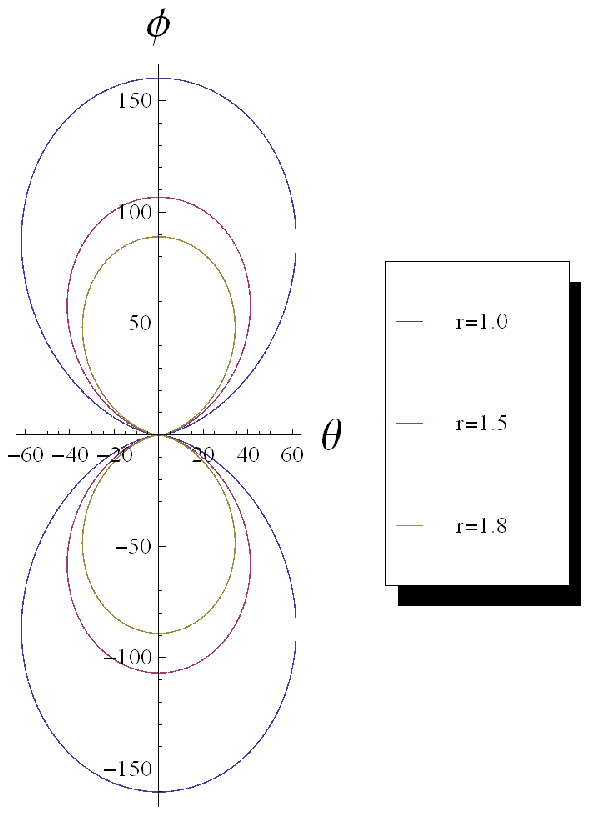}
\caption{\label{fig:AmplAbl} The angular part of amplitude $P_{A^{a}_{+}}(\theta, r)$ (\ref{eq:PAtimes+A}) pictured in dependence on $\theta$ angle and additional $\phi$ angle in radians at the time $t=0$ [s] in 3D and 2D figures. The amplitude has a shape of toroid with symmetry around axes $z=0$. The dependence on $1/r$ is depicted in smaller surfaces in the figure, the biggest surface is $r=1$ m, then $r=1.5$ m and $1.8$ m. The surface is getting smaller as $r\rightarrow 10$ m  (at the distance of the detector) and approaches $0$ as $r\rightarrow \infty$. The toroid was cut on purpose to see the inner surfaces of lower $r$.
The polar 2D diagram was plotted for fixed angle $\phi=\pi/2$.}
\end{figure}

\begin{figure}[h]
\centering
\includegraphics[width=0.5\textwidth]{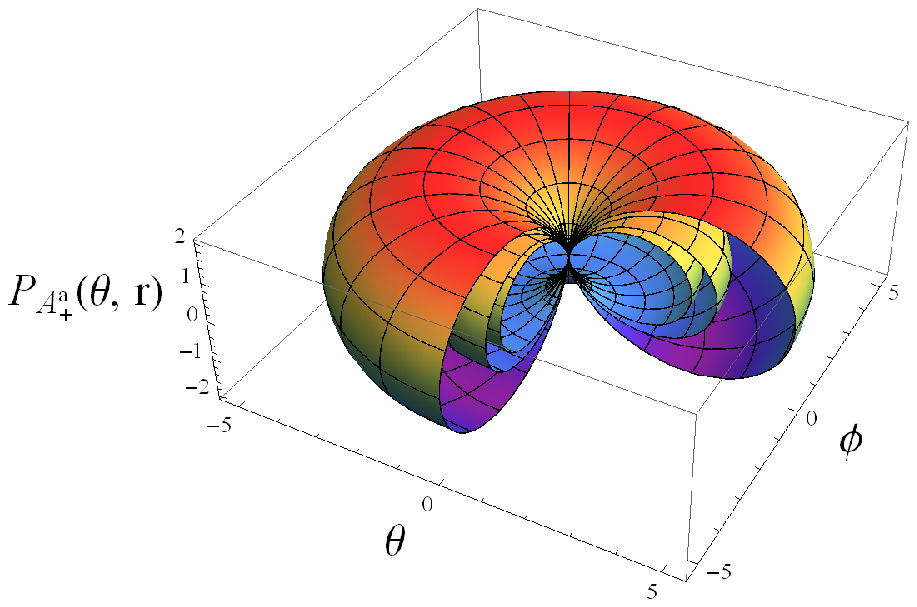}
\includegraphics[width=0.4\textwidth]{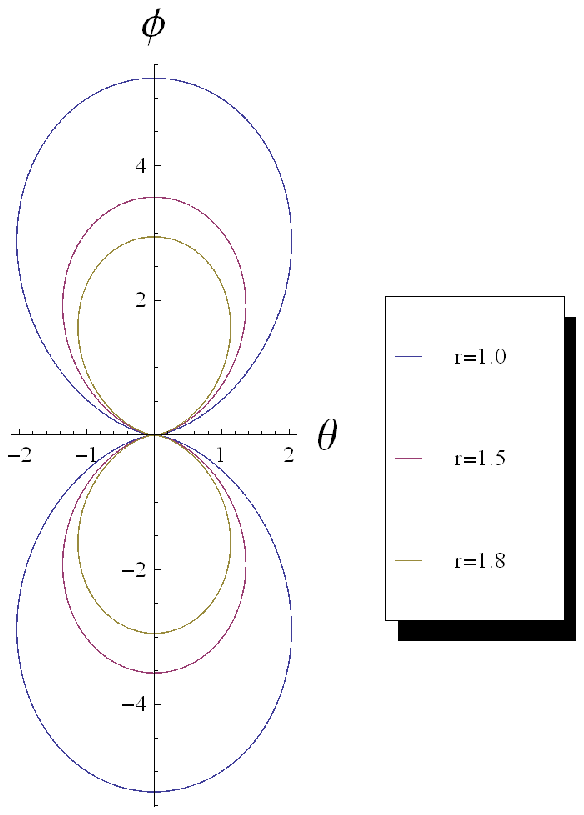}
\caption{\label{fig:AmplAbl1} The angular part of amplitude $P_{A^{a}_{+}}(\theta, r)$ (\ref{eq:PAtimes+A}) pictured in dependence on $\theta$ angle and additional $\phi$ angle in radians at the time $t=8\mu$s in 3D and 2D figures. The amplitude has a shape of toroid with symmetry around axes $z=0$. The dependence on $1/r$ is depicted in smaller surfaces in the figure, the biggest surface is $r=1$ m, then $r=1.5$ m and $1.8$ m.  The toroid was cut on purpose to see the inner surfaces of lower $r$.}
\end{figure}

To visualize the amplitudes it is convenient to rewrite them as
\begin{align}
A^{a}_{+}(t;\theta,\phi)&=4\frac{G}{c^4}S\rho_{0}v^2_{r}P_{A^{a}_{+}}(\theta),\nonumber\\
A^{a}_{\times}(t;\theta,\phi)&=-4\frac{G}{c^4}S\rho_{0}v^2_{r}P_{A^{a}_{\times}}(\theta),\label{eq:hkn2}
\end{align}
where the angular dependence is denoted as
\begin{align}
P_{A^{a}_{+}}(\theta,r)&=\frac{1}{r}\left\{(\frac{1}{2}+b_{I}e^{-b_{I}})\sin{2\theta}(a\sin\phi+b\cos\phi)\right.\\
&-\left.3\sin^2\theta\left[3(-v_{r}t+f_{2})(-\frac{2}{3}+e^{-b_{I}}(b^2_{I}+1))\right.\right.\nonumber\\
&\left.\left.-2z_{L}e^{-b_{I}}(3-\frac{1}{\dot{z}_{r}})\right]\right\},\label{eq:PAtimes+A} \\
P_{A^{a}_{\times}}(\theta,r)&=\frac{1}{r}(\frac{1}{2}+b_{I}e^{-b_{I}})\sin\theta(a\cos\phi-b\sin\phi).\label{eq:PAtimesxA}
\end{align}
We have included the $r$ dependence in the angular parts of the amplitudes in order to investigate the dependence.
Let us note that the time when the radiation reaches the detector is
\begin{align}
t_{det}=f_{2}/v_{r},\label{eq:tdetectorA}
\end{align}
then the geometrical structure of $P_{A^{a}_{+}}(\theta,r)$ changes because of $f_{2}-v_{r}t_{det}=0$. The choice  of coordinates enables us to choose $f_{2}$, this change of structure is then just of coordinate nature and has no physical meaning.
We have plotted the amplitude $A^{a}_{+}$ in the following graphs Fig.~\ref{fig:AmplAbl} and Fig.~\ref{fig:AmplAbl1}. The graphs were made for values $a=b=1\,{\rm mm}=0.1\,{\rm cm}$,  $I_{L}=50\,[\rm PW/{\rm cm}^2]$ and  $R_{t}=15.144[{\rm kg}/{\rm m}^3]$  for Carbon. The velocity $v_{r}=1.14\times 10^6 {\rm [m/s]}$ and $b_{I}=0.2$ starts at this value as is growing in time. The amplitude $A_{A^{a}_{+}}=4.34\times 10^{-41}$ and $A_{A^{a}_{\times}}=-4.34\times 10^{-41}$.

The angular shape of $P_{A^{a}_{+}}(\theta,t)$ of the ablation wave at start $t=0$ is depicted in Fig.~\ref{fig:AmplAbl}. The angular dependence has a symmetric shape of toroid with the center at $z=0$ ($\theta=\phi=0$). The surfaces inside the toroid represent angular structure for larger $r$ and we observe that the magnitude of the toroid becomes smaller as expected as $1/r$. Before tha radiation reaches the detector $t<t_{det}$, $t=8\mu$s, the amplitude is smaller Fig.~\ref{fig:AmplAbl1} than Fig.~\ref{fig:AmplAbl}.

\begin{figure}[h]
\centering
\includegraphics[width=0.48\textwidth]{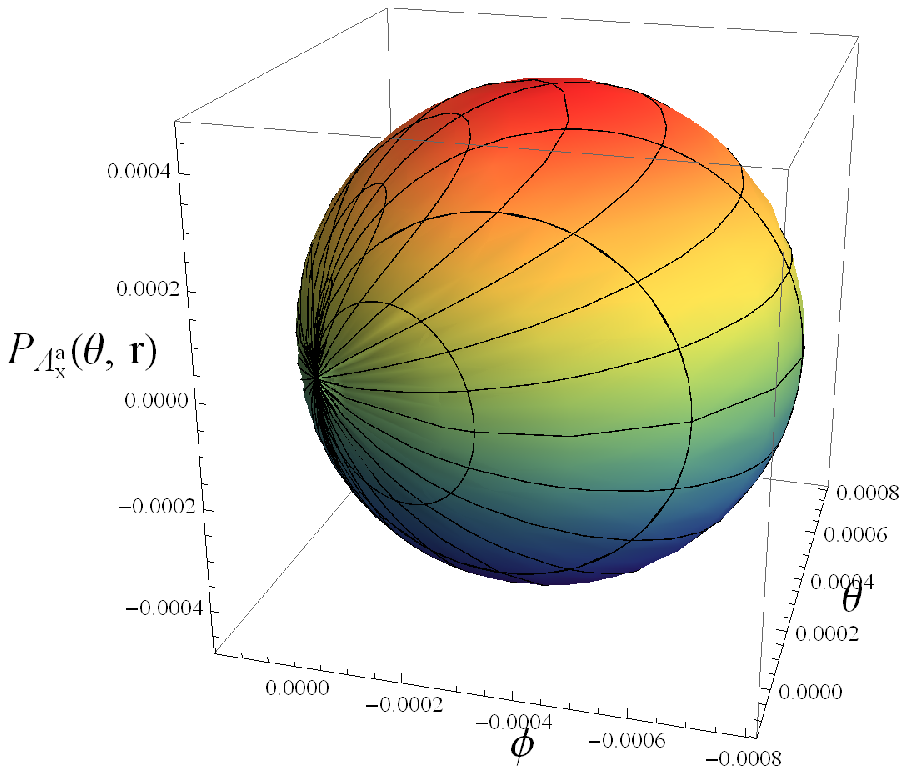}
\includegraphics[width=0.48\textwidth]{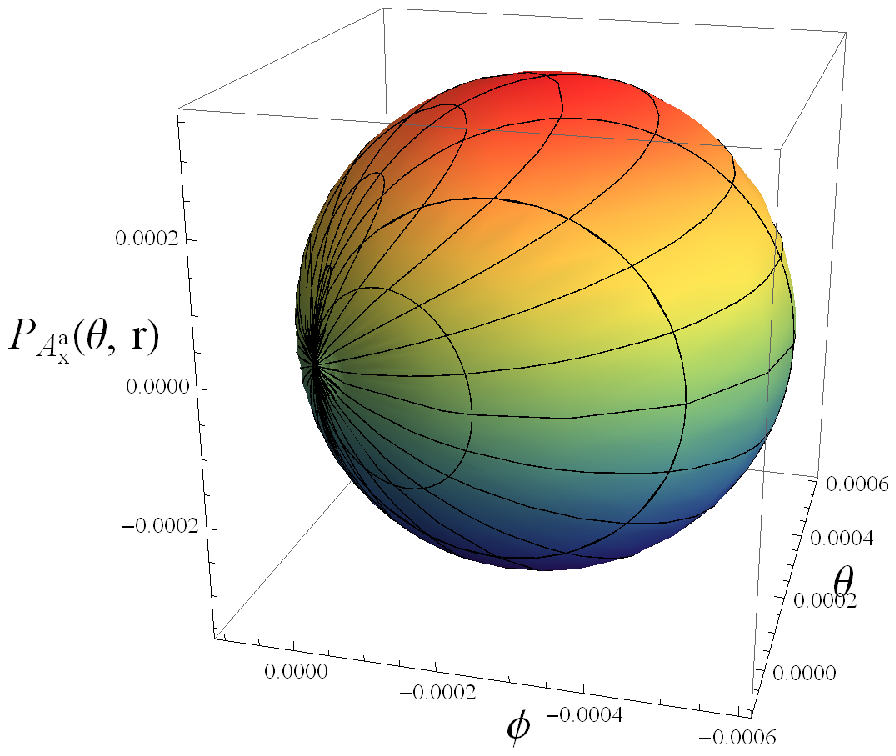}
\caption{\label{fig:AmplAbl2} The angular part of amplitude $P_{A^{a}_{\times}}(\theta, r)$ (\ref{eq:PAtimesxA}) pictured in dependence on $\theta$ angle and additional $\phi$ angle in radians at the time $t=0$ in 3D. The amplitude has a shape of a ball with start at $z=0$.}
\end{figure}

The image of the amplitude $A^{a}_{\times}$ in depicted in the Fig.~\ref{fig:AmplAbl2}, where the first image is for $t=0$ and the second one for $t=8\mu$s. The amplitude is slightly decreasing in time as the previous $A^{a}_{x}$ amplitude.

The orientation of the both amplitudes on left toward each other are very similar to ones for the shock wave model, see Fig.~8 in \cite{Kadlecova2015}, therefore we will not present them again. 

The difference in the time dependency of the two independent polarization modes might be very important for the experimetal detection, because it would be possible to distinguish the two modes of polarization.

\subsubsection{The case of piston model}

Afterwards we use the mass moments expressed in terms of derivatives of function $z$, the amplitudes read as follows,
\begin{align}
A^{p}_{+}(t;\theta,\phi)&=\frac{1}{r}\frac{G}{c^4}S\rho_{0}\left[-\frac{1}{3}(z_{p}^3)^{\ddot{}}\sin^2\theta\right.\nonumber\\
&+\left.\frac{1}{4}(z_{p}^2)^{\ddot{}}\sin(2\theta)(a\sin\phi+b\cos\phi)\right.\nonumber\\
&+\left.\frac{1}{3}{\ddot{z}_{p}}a^2(\cos^2\phi-\sin^2\phi\cos^2\theta)\right.\nonumber\\
+&\left.\frac{1}{3}{\ddot{z}_{p}}b^2(\sin^2\phi-\cos^2\phi\cos^2\theta)\right.\nonumber\\
&-\left.\frac{3}{4}S{\ddot{z}_{p}}\sin(2\phi)(1+\cos^2\theta)\right],\label{eq:hpn1}\\
A^{p}_{\times}(t;\theta,\phi)&=\frac{1}{r}\frac{G}{c^4}S\rho_{0}\left[-\frac{1}{2}(z_{p}^2)^{\ddot{}}\sin\theta(a\cos\phi-b\sin\phi)\right.\nonumber\\
&+\left.\ddot{z}_{p}\cos\theta\left(\frac{1}{3}(a^2-b^2)\sin{2\phi}+\frac{1}{2}S\cos{2\phi}\right)\right].\label{eq:hkn1}
\end{align}
After we use the ansatz for the $z_{p}$, we get
\begin{align}
A^{p}_{+}(t;\theta,\phi)&=\frac{1}{2r}\frac{G}{c^4}S\rho_{0}{v^{2}_{p}}\left[-4v_{p}t\sin^2\theta+\sin{2\theta}(a\sin\phi+b\cos\phi)\right], \label{eq:hpn2p}\\
A^{p}_{\times}(t;\theta,\phi)&=-\frac{1}{r}\frac{G}{c^4}S\rho_{0}{v_{p}^{2}}\sin\theta(a\cos\phi-b\sin\phi)\label{eq:hkn2p}.
\end{align}
The final expressions (\ref{eq:hpn2p}) and (\ref{eq:hkn2p}) are very similar to results in Eq.~(66--67) \cite{Kadlecova2015}. The difference is in the positive sign of the second term in (\ref{eq:hpn2p}) and minus sign in the whole expression (\ref{eq:hkn2p}).
We will rewrite the amplitudes into
\begin{align}
A^{p}_{+}(t;\theta,\phi)&=\frac{1}{2}\frac{G}{c^4}S\rho_{0}{v^{2}_{p}}P_{A^{p}_{+}}, \label{eq:hpn2}\\
A^{p}_{\times}(t;\theta,\phi)&=-\frac{G}{c^4}S\rho_{0}{v_{p}^{2}}P_{A^{p}_{\times}}\label{eq:hkn2},
\end{align}
where we denote the angular part of the amplitude
\begin{align}
P_{A^{p}_{+}}(\theta,r)&=\frac{1}{r}\left[-4v_{p}t\sin^2\theta+\sin{2\theta}(a\sin\phi+b\cos\phi)\right], \label{eq:hpn2pa+}\\
P_{A^{p}_{\times}}(\theta,r)&=\frac{1}{r}\sin\theta(a\cos\phi-b\sin\phi)\label{eq:hkn2pax}.
\end{align}

The graphs were made with the parameters, $r=10$ m, parameters $a, b$ of the target foil are $a=b=1\,{\rm \mu m}=1\times 10^{-6}\,{\rm m}$ and therefore $I_{L}=7\times 10^{8}\,[\rm PW/{\rm cm}^2]$ and the velocity $v_{p}=153008\,{\rm [km/s]}$.

In the following figures, we will observe the effect of time dependence of the $A^{p}_{+}$ amplitude. The angular shape of $P_{A^{p}_{+}}(\theta,t)$ of the piston at start $t=0$ is depicted in Fig.~\ref{fig:Ampl++}. The angular dependence has a symmetric shape of cloverleaf with the center at $z=0$ ($\theta=\phi=0$), because the first term in (\ref{eq:hpn2pa+}) vanishes. The surfaces inside the cloverleaf represent angular structure for larger $r$ and we observe that the magnitude of the cloverleaf becomes smaller as expected as $1/r$. For shock wave model, we got this geometry structure for the detection time and the shape was of coordinate nature -- choice of the start of coordinates. The reason we obtain the geometry here is because we have chosen the start of coordinates in the opposite way than the shock wave model set up, therefore we get the structure at the start of the experiment. 

At the time shortly before the detector $t<t_{det}$, the angular dependence is larger in Fig.~\ref{fig:Ampl++Detector} than the one at $t=0$ in the previous Fig.~\ref{fig:Ampl++} and the geometry changes to the toroidal geometry as in the shock wave model. Then the time when radiation reaches detector is $t_{det}=1.3\times 10^{-6}$s and the amplitudes $A_{A^{p}_{+}}=9.7\times 10^{-38}$ and $A_{A^{p}_{\times}}=1.95\times 10^{-37}$.

\begin{figure}[h]
\centering
\includegraphics[width=0.4\textwidth]{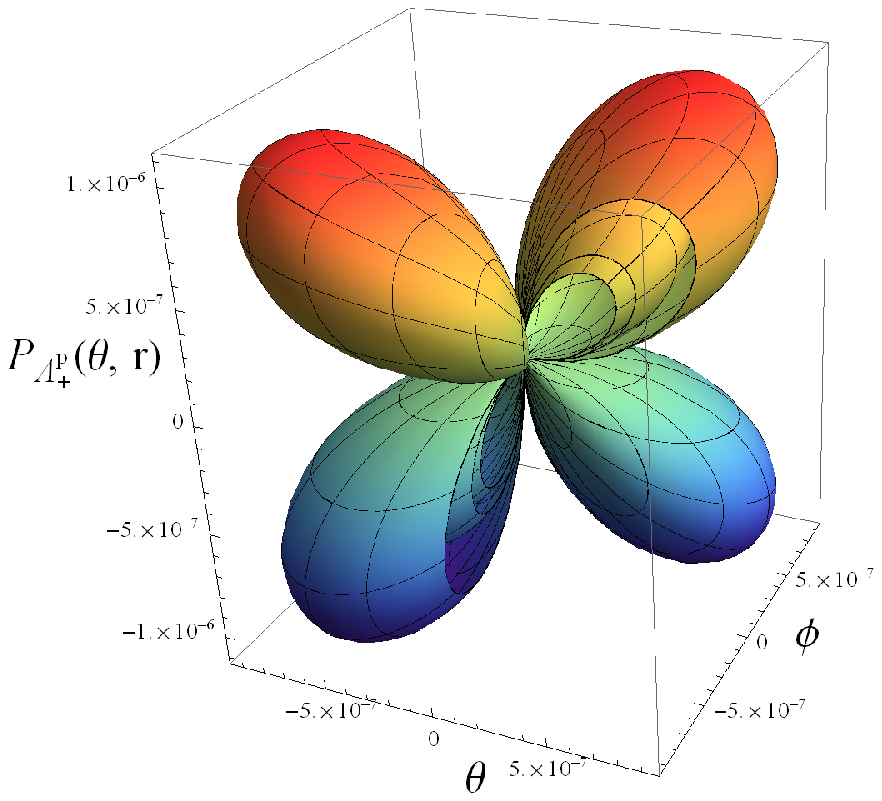}
\includegraphics[width=0.3\textwidth]{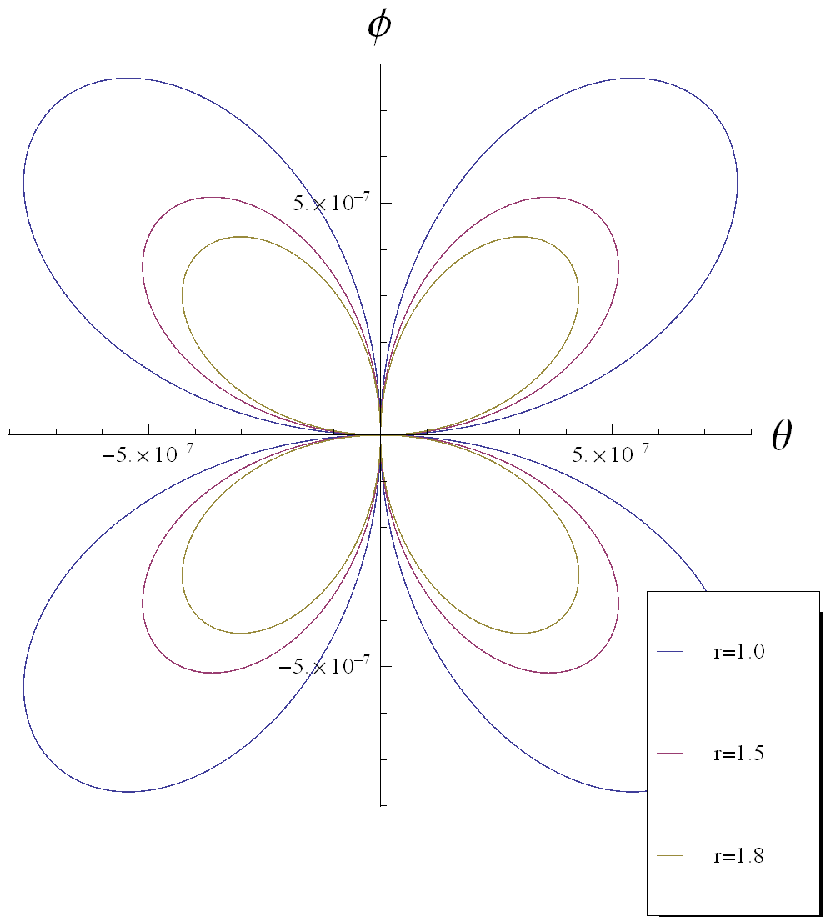}
\caption{\label{fig:Ampl++} The angular part of amplitude $P_{A^{p}_{+}}(\theta, r)$ (\ref{eq:hpn2pa+}) pictured in dependence on $\theta$ angle and additional $\phi$ angle at the time $t=0$ s in 3D and 2D figures. The geometry has a shape of cloverleaf with symmetry around $z=0$. The dependence on $1/r$ is depicted in smaller surfaces in the figure, the biggest surface is $r=1$ m, then $r=1.5$ m and $1.8$ m. The surface is getting smaller as $r\rightarrow 10$ m (at the distance of the detector) and approaches $0$ as $r\rightarrow \infty$.  The polar 2D diagram was plotted for fixed angle $\phi=\pi/2$. The left image was cut out on purpose to see the inner surfaces.}
\end{figure}

\begin{figure}[h!]
\centering
\includegraphics[width=0.4\textwidth]{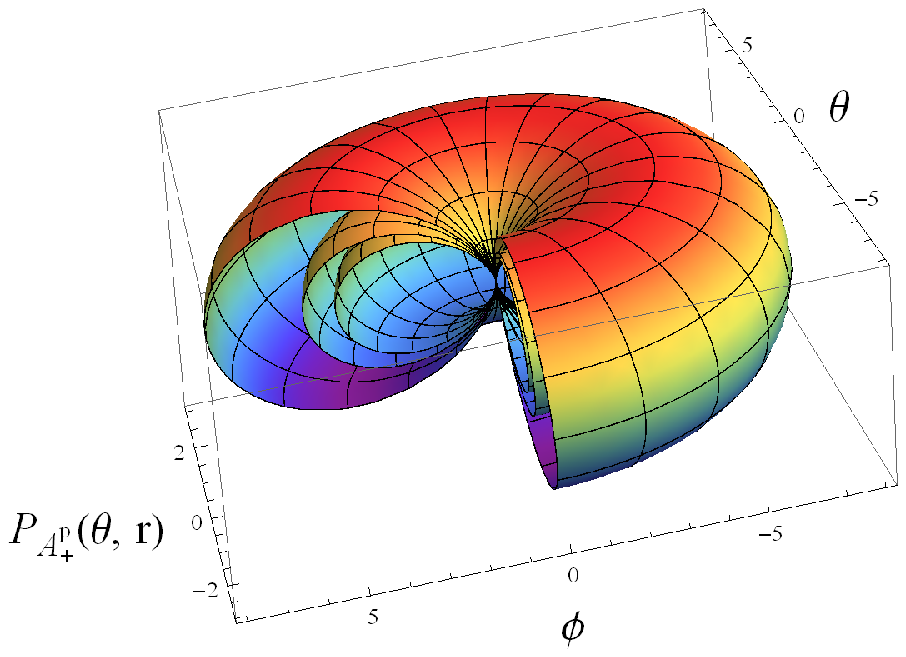}
\includegraphics[width=0.3\textwidth]{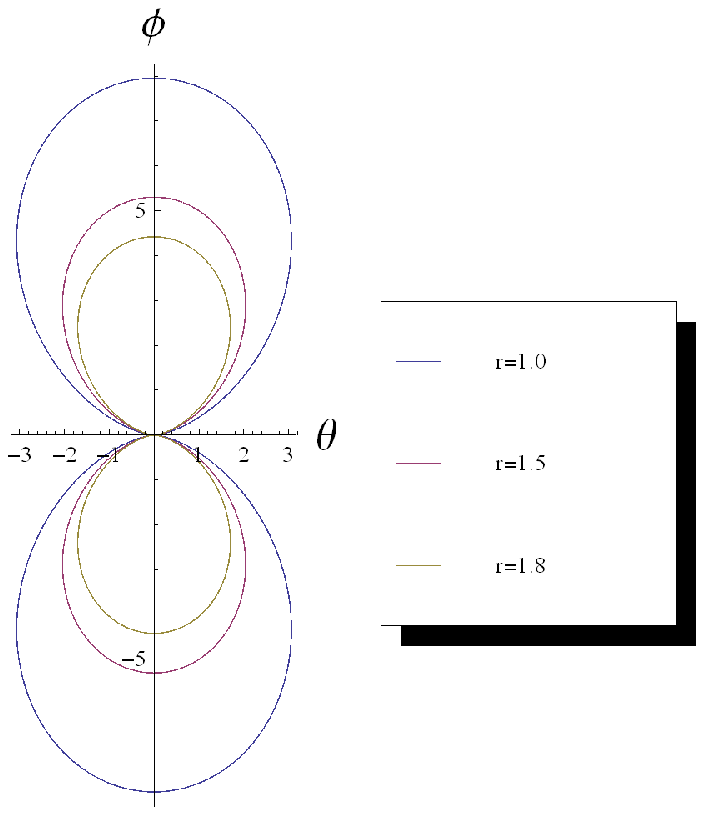}
\caption{\label{fig:Ampl++Detector} The angular part of amplitude $P_{A^{p}_{+}}(\theta, r)$ (\ref{eq:hpn2pa+}) pictured in dependence on $\theta$ angle and additional $\phi$ angle at the time $t=130\,\mu$s in 3D and 2D figures. The dependence on $1/r$ is depicted in smaller surfaces in the figure, the biggest surface is $r=1$ m, then $r=1.5$ m and $1.8$ m. The surface is getting smaller as $r\rightarrow 10$ m (at the distance of the detector) and approaches $0$ as $r\rightarrow \infty$.
The polar 2D diagram was plotted for fixed angle $\phi=\pi/2$.}
\end{figure} 

\begin{figure}[h!]
\centering
\includegraphics[width=0.4\textwidth]{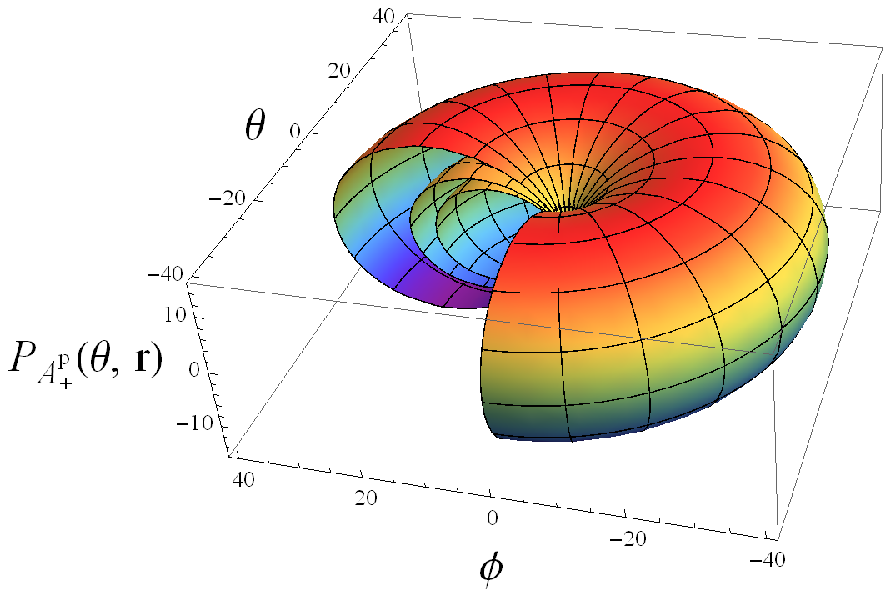}
\includegraphics[width=0.3\textwidth]{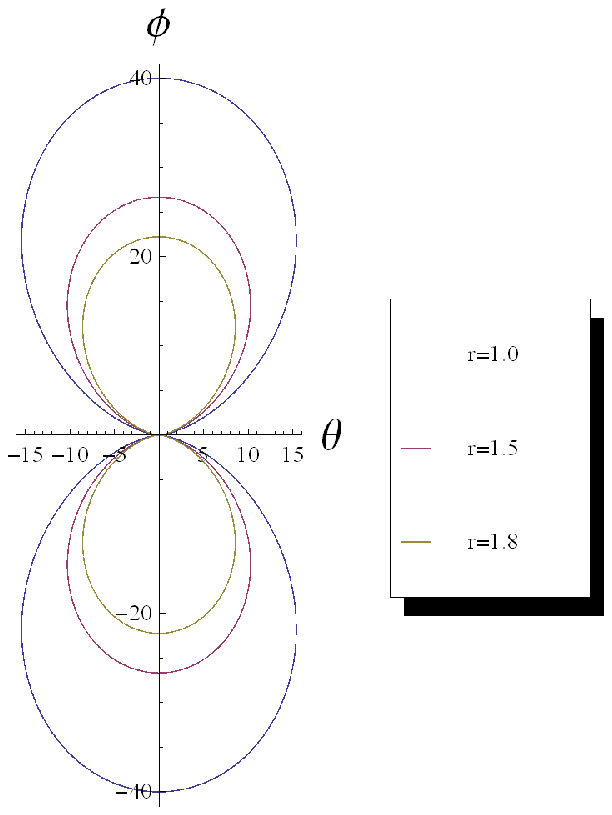}
\caption{\label{fig:AmplDetector} The angular part of amplitude $P_{A^{p}_{+}}(\theta, r)$ (\ref{eq:hpn2pa+}) pictured in dependence on $\theta$ angle and additional $\phi$ angle in radians at the detector in 3D and 2D figures. The magnitude of the angular part of amplitude is much smaller than previous ones. 
The polar 2D diagram was plotted for fixed angle $\phi=\pi/2$.}
\end{figure}

At the moment $t_{det}$ when the radiation reaches the detector, the geometry does not change in Fig.~\ref{fig:AmplDetector}, we can see the structure of toroid again.  We observe that the amplitude of the angular dependence is much larger than the two previously pictured.

The amplitude for polarization mode $\times$ is the almost identical to Fig.~6 in \cite{Kadlecova2015} up to amplitude (\ref{eq:hkn2}) which has opposite sign. Also the orientation of both amplitudes $x$ and $\times$ is similar to Fig.~8 on the left in \cite{Kadlecova2015}. The toroidal amplitudes are rotated for 180$^\circ$ in $\theta$ compared to the images for shock wave model, the Figs. \ref{fig:Ampl++Detector} and \ref{fig:AmplDetector} are rotated accordingly to show off the inside layers.

The difference in the time dependency of the two independent polarization modes might be very important for the experimetal detection in both shock wave and piston models in the quadrupole approximation of linear gravity.

\subsection{The radiative characteristics for generated gravitational waves}
In this part, we will calculate radiative characteristics along the expressions in subsection (4.3) in chapter IV \cite{Kadlecova2015}. 
First, we will concentrate on ablation model and then on piston model.

\subsubsection{The case of ablation model}
In our case, the amplitudes are time--independent for ablation model, then the invariant density Eq.~(71) in \cite{Kadlecova2015} reads,
\begin{align}
t_{00}^{GW}=\frac{81}{\pi}\frac{G}{r^2c^4}S^2\rho_{0}^2v_{r}^6\sin^4\theta(-\frac{2}{3}+e^{-b_{I}}(b^2_{I}+1))^2,
\end{align}
which functionally depends on $r$ and $\theta$ angle. The energy goes to zero as $r$ approaches infinity, where we have neglected terms of type $e^{-b_{I}}/z_{r}$.

The energy spectrum is then trivial $\frac{d E}{d A}=\frac{c^3}{16\pi G}\int_{0}^{\tau}d t (\dot{A^{a}_{+}}^2+\dot{A^{a}_{\times}}^2)=\frac{9}{4\pi r c}S\rho_{0}v^3_{r}\sin^2\theta\int_{0}^{\tau}(-\frac{2}{3}+e^{-b_{I}}(b^2_{I}+1))$ where $d A=r^2 d \Omega$ is surface element and $\tau$ is duration of pulse.

Now, we are able to substitute the ansatz for the $z_{s}$ into into Eqs.~(74--76) in \cite{Kadlecova2015}. Then the expressions for the wave vector in directions $n=x, y, z$ read,
 \begin{align}
\mathcal{S}^{a}_{n_{x}}&=\mathcal{S}^{a}_{n_{y}}=\frac{11}{12}\frac{c^3}{G\pi r^2}S^2{\rho_{0}^2}{v_{r}^{6}}\left(-\frac{2}{3}+e^{-b_{I}}(b^2_{I}+1)\right)^2,\nonumber\\
\mathcal{S}^{a}_{n_{z}}&=-\frac{16}{3}\frac{c^3}{G\pi r^2}S^2{\rho_{0}^2}{v_{r}^{6}}\left(-\frac{2}{3}+e^{-b_{I}}(b^2_{I}+1)\right)^2,\label{eq:snxyz}
\end{align}
and the expression for the general wave vector  Eq.~(77) in \cite{Kadlecova2015} results in
 \begin{align}
\mathcal{S}^{a}_{n}=&\frac{16}{9}\frac{S^2c^3{\rho_{0}^2}{v_{r}^{6}}}{G\pi r^2}\left(-\frac{2}{3}+e^{-b_{I}}(b^2_{I}+1)\right)^2\nonumber\\
&\times\left[9-(\sin^2{\theta}+16\cos^{2}\theta)+(2\cos^{2}\theta+\sin^{2}\theta)^2\right]\label{eq:radCharSnAblation}.
\end{align}

The radiative characteristics (\ref{eq:radCharSnAblation}) depends only on the $\theta$ angle which is a consequence of the axis symmetry of the problem. The characteristics behave as $1/r^2$ as ~$r\rightarrow\infty$ contrary to $1/r$ decay of amplitudes.
To visualize the characteristic, it is useful to separate the angular part from its amplitude as
\begin{align}
\mathcal{S}^{a}_{n}=&\frac{16}{9}\frac{S^2c^3{\rho_{0}^2}{v_{r}^{6}}}{G\pi}\left(-\frac{2}{3}+e^{-b_{I}}(b^2_{I}+1)\right)^2P_{S^{a}_n}(\theta)\label{eq:radCharSnPAblation}
\end{align}
where the angular dependence
\begin{align}
P_{S^{a}_n}(\theta)= \frac{1}{r^2}\left[9-(\sin^2{\theta}+16\cos^{2}\theta)+(2\cos^{2}\theta+\sin^{2}\theta)^2\right].\label{fig:PAblation}
\end{align}
In the calculations we have neglected the terms of type $e^{-b_{I}}/z_{r}$ as in previous calculations.

\begin{figure}[h!]
\centering
\includegraphics[width=0.45\textwidth]{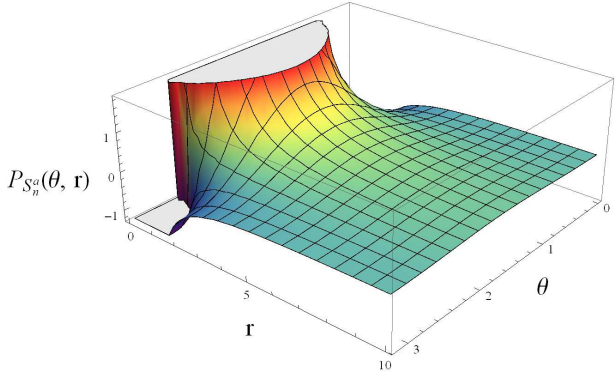}
\caption{\label{fig:CharRadAbl} The radiation characteristics $\mathcal{S}^{a}_{n}$ (\ref{eq:radCharSnPAblation}) pictured in dependence on $\theta$ angle and $r$. We have plotted just the angular dependence $P_{S^{a}_{n}}(\theta)$ (\ref{fig:PAblation}), $\mathcal{S}^{a}_{n}=A_{S^{a}_n}P_{S^{a}_{n}}(\theta)$, (\ref{eq:radCharSnPAblation}). The surface is approching $0$ at the distance of the detector $r=10$ m, also while $r\rightarrow \infty$ the surface approaches zero.}
\end{figure}

\begin{figure}[h!]
\centering
\includegraphics[width=0.45\textwidth]{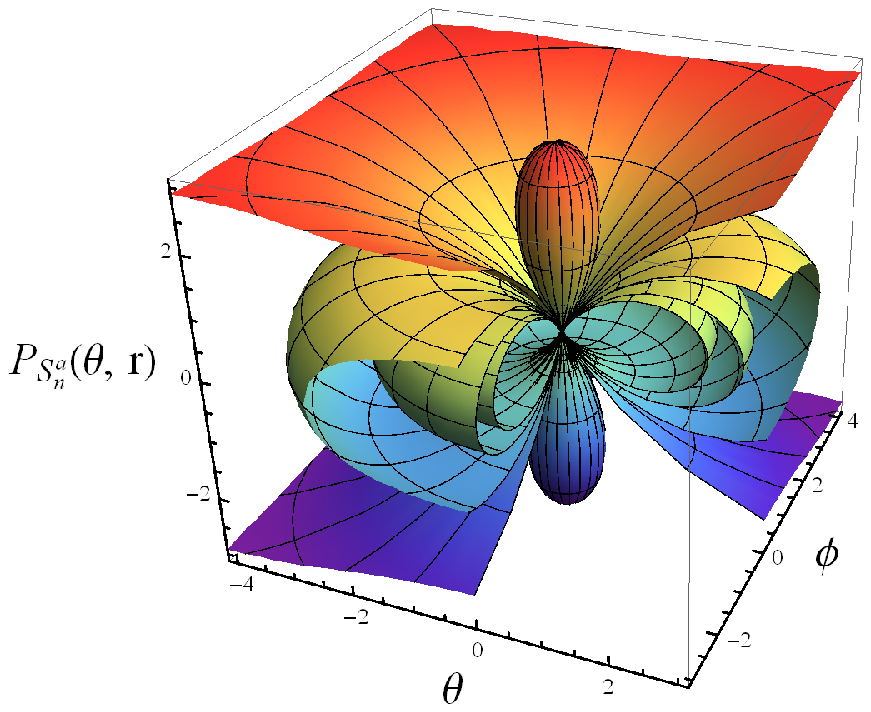}\\
\includegraphics[width=0.45\textwidth]{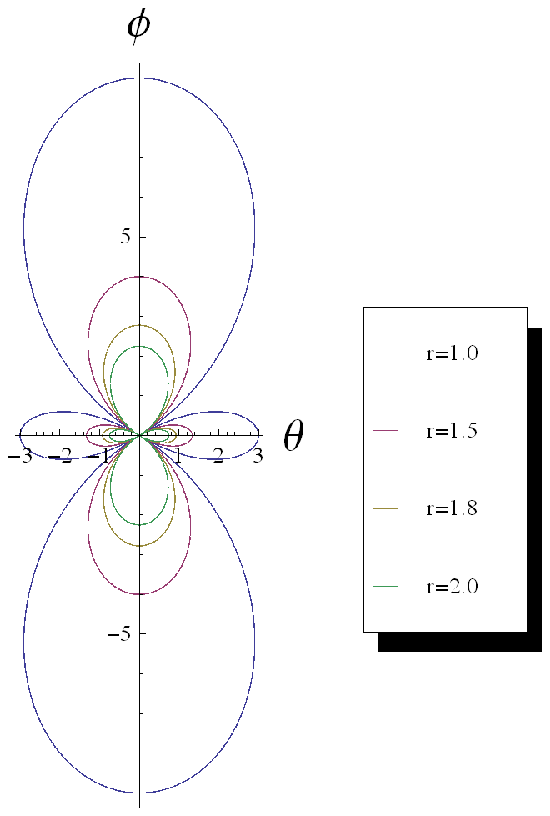}
\caption{\label{fig:CharRadR} The radiation characteristics $\mathcal{S}^{a}_{n}$ (\ref{eq:radCharSnPAblation}) pictured in dependence on $\theta$ angle and $r$ and rotated additionally around $\phi$ angle in radians. We have plotted just the angular dependence $P_{S^{a}_{n}}(\theta)$ (\ref{fig:PAblation}), $\mathcal{S}^{a}_{n}=A_{S^{a}_n}P_{S^{a}_{n}}(\theta)$, (\ref{eq:radCharSnPAblation}). The dependence on $1/r^2$ is depicted in smaller surfaces in the figure, the biggest surface is $r=1$ m, then $r=1.5$ m, $1.8$ m and $r=2$ m. The surface is getting smaller as $r\rightarrow 10$ m(the distance of the detector), while $r\rightarrow \infty$ the surface approaches $0$. The structure of surfaces is symmetric around the axes $z=0$. The image on the left is partially cut out to see the inner surfaces.}
\end{figure}

The radiation structure is pictured in Fig.~\ref{fig:CharRadAbl}.
The amplitude $A_{S^{a}_n}$ (\ref{eq:radCharSnPAblation}) has a specific value $A_{S^{a}_n}=8.94\times 10^{64}$, for values $a=b=1\,{\rm mm}=0.1\,{\rm cm}$,  $I_{L}=50\,[\rm PW/{\rm cm}^2]$, where the velocity $v_{r}=1.14\times 10^6 {\rm [m/s]}$. The dependence on $\theta$ and $r$ is plotted in Fig.~\ref{fig:CharRadAbl} and the polar dependence on $\theta,\,\phi$ and $r$ is plotted in Fig.~\ref{fig:CharRadR} (2D and 3D).
 
This directional characteristic would help with the experimental set up and positions of the detectors. The directional structure (\ref{fig:CharRadAbl}) has similar toroidal shape as the structure for shock wave model Fig.~9 in \cite{Kadlecova2015} and piston model (\ref{fig:CharRadPolar}) but with the additional radiative part in the  $z=0$ direction of shape of a dumbbell. It suggests existence of longitudinal GW radiation in the direction of the laser propagation, which should not occur in linear gravity, and is the consequence of the broken mass conservation law as mentioned earlier. 

\subsubsection{The case of piston model}
Again, the $A_{\times}$ amplitude is time--independent, therefore just the $A^{p}_{+}$ contributes to the effective tenzor,
\begin{align}
t_{00}^{GW}=\frac{c^4}{32\pi G}\langle \dot{A^{p}_{+}}^2\rangle=\frac{1}{8\pi}\frac{G}{r^2c^4}S^2\rho_{0}^2v_{p}^6\sin^4\theta,
\end{align}
which functionaly depends on $r$ and $\theta$ angle. The energy goes to zero as $r$ approaches infinity. The energy spectrum is then trivial $\frac{d E}{d A}=\frac{c^3}{16\pi G}\dot{A}^2_{+}\tau$.

Now, we are able to substitute the ansatz for the $z_{s}$ into Eqs.~(74--76) in \cite{Kadlecova2015}. Then the expressions for the wave vector in directions $n=x, y, z$ read,
 \begin{align}
\mathcal{S}^{p}_{n_{x}}=\mathcal{S}^{p}_{n_{y}}=&\frac{c^3}{36G\pi r^2}S^2{\rho_{0}^2}{v_{p}^{6}},\;
\mathcal{S}^{p}_{n_{z}}=\,0,\nonumber
\end{align}
and the expression for the general wave vector Eq.~(120) in \cite{Kadlecova2015} results in
 \begin{align}
\mathcal{S}^{p}_{n}=&\frac{S^2c^3{\rho_{0}^2}{v_{p}^{6}}}{324G\pi r^2}\left[12-4(\sin^2{\theta}+4\cos^{2}\theta)+(2\cos^{2}\theta-\sin^{2}\theta)^2\right]\label{eq:radCharSn}.
\end{align}

The radiative characteristics (\ref{eq:radCharSn}) depends only on the $\theta$ angle which is a consequence of the axis symmetry of the problem. The characteristics behave as $1/r^2$ as~$r\rightarrow\infty$ contrary to $1/r$ decay of amplitudes.
To visualize the characteristic, it is useful to separate the angular part from its amplitude as
\begin{align}
\mathcal{S}^{p}_{n}=&\frac{S^2c^3{\rho_{0}^2}{v_{p}^{6}}}{324G\pi}P_{S^{p}_{n}}(\theta)\label{eq:radCharSnP}
\end{align}
where the angular dependence
\begin{align}
P_{S^{p}_n}(\theta)= \frac{1}{r^2}\left[12-4(\sin^2{\theta}+4\cos^{2}\theta)+(2\cos^{2}\theta-\sin^{2}\theta)^2\right].\label{fig:P}
\end{align}

\begin{figure}[h!]
\centering
\includegraphics[width=0.45\textwidth]{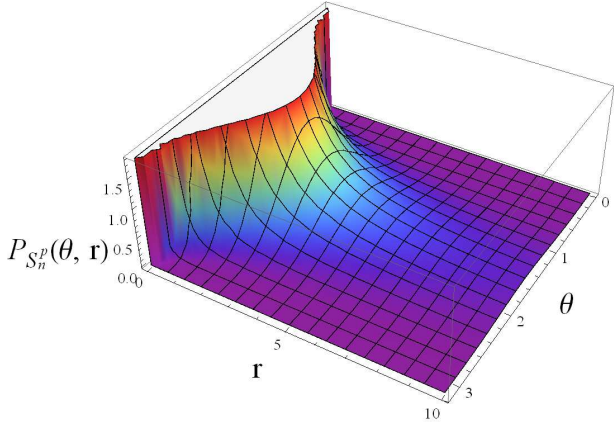}
\caption{\label{fig:CharRad} We have plotted just the angular dependence $P_{S^{p}_{n}}(\theta)$ (\ref{fig:P}), $\mathcal{S}^{p}_{n}=A_{S^p_n}P_{S^p_{n}}(\theta)$, (\ref{eq:radCharSnP}) in dependence on angle $\theta$ in radians and distance $r$ in meters.}
\end{figure}

\begin{figure}[h!]
\centering
\includegraphics[width=0.45\textwidth]{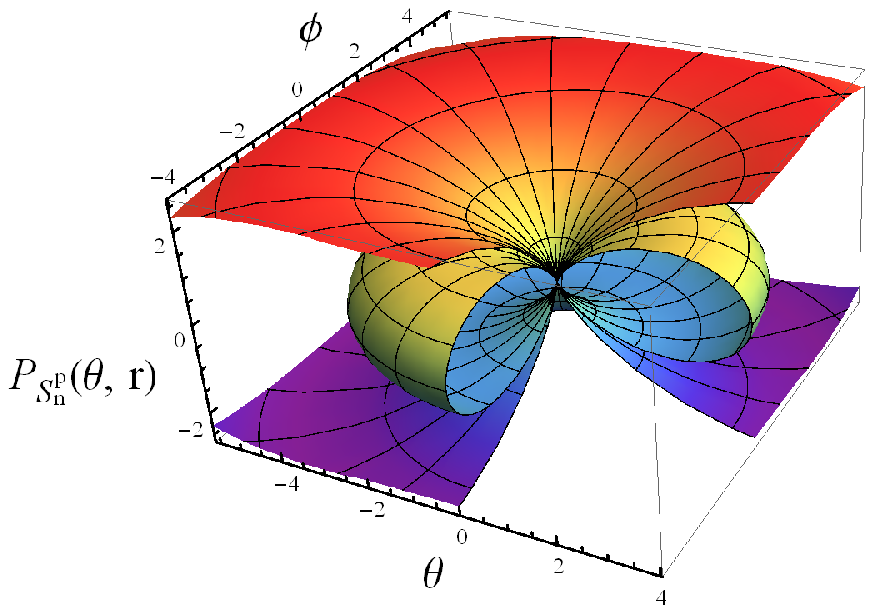}\\
\includegraphics[width=0.45\textwidth]{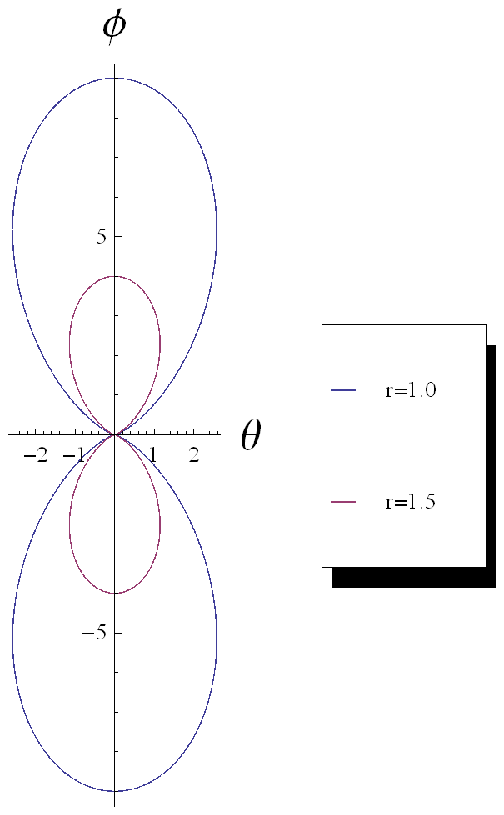}
\caption{\label{fig:CharRadPolar}The radiation characteristics $\mathcal{S}^{p}_{n}$ (\ref{eq:radCharSn}) pictured in dependence on $\theta$ angle and rotated additionally around $\phi$ angle in radians. We have plotted just the angular dependence $P_{S^{p}_{n}}(\theta)$ (\ref{fig:P}), $\mathcal{S}^{p}_{n}=A_{S^p_n}P_{S^p_{n}}(\theta)$, (\ref{eq:radCharSnP}). The dependence on $1/r^2$ is depicted in smaller surfaces in the figure, the biggest surface is $r=1$ m, then $r=1.5$ m. The surface is getting smaller as $r\rightarrow 10$ m (the distance of the detector), while $r\rightarrow \infty$ the surface approaches $0$. The structure of surfaces is symmetric around the axes $z=0$. The image on the left is cut out on purpose to see the inner structure.}
\end{figure}

The radiation structure is pictured in Fig.~\ref{fig:CharRadPolar} which is the same as for the shock wave model thanks to the same resulting formulae (\ref{fig:P}).
The amplitude $A_{S^p_n}$ (\ref{eq:radCharSnP}) has a specific value $A_{S^p_n}=4.54\times 10^{69}$ for $a=b=10^{-6}$m, $I_{L}=7\times 10^{8}$ PW${/\rm m^2}$ and $v_{s}=153008$ m$/$s. The polar dependence on $\theta,\,\phi$ and $r$ in Fig.~\ref{fig:CharRad} and the dependence on $\theta$ and $r$ is plotted in Fig.~\ref{fig:CharRadPolar}.
 
The directional structure of radiation is the same for the shock wave model and for the piston model in the approximation we use in the paper, etc. the gravity in linear approximation up to quadrupole moment in the moment expansion. The differences might appear in higher orders of the expansion.

\subsection{The angular momentum}
The angular momentum carried away per unit time by the gravitational waves is given by Eq.~(81) in \cite{Kadlecova2015}, we obtain for ablation and piston model (using derivatives in (\ref{A3}))
\begin{align}
\left(\frac{d J_{ablation}^{i}}{dt}\right)_{quad}&=0 \longrightarrow  J^{i}_{ablation}=\text{const},\label{eq:angul0}\\
\left(\frac{d J_{piston}^{i}}{dt}\right)_{quad}&=0 \longrightarrow  J^{i}_{piston}=\text{const},\label{eq:angul1}
\end{align}
and the angular momentum of the radiation in the shock wave model stays constant in time due to the single dimension of the experiment. In case of ablation model, we have neglected the terms of type $e^{-b_{I}}/z_{r}$ to obtain the result.

\section{\label{sec:number5}The behaviour of test particles in the presence of gravitational wave}
We will analyse the test particles for the ablation and piston models in the same way as section V in \cite{Kadlecova2015}.

\subsection{The predictions for detector}
According to Eq.~(83) in \cite{Kadlecova2015}, we can estimate the linear size $L$ of the possible detector
\begin{align}
L_{ablation} &\ll 4.7746\cdot 10^{-2}\, {\rm m},\quad
L_{piston} \ll 4.778\times 10^{-5}\, {\rm m},\label{eq:DetectLambdaPiston}
\end{align}
which might serve as usefull estimation for validity of the future experiment and the detector. We have used the numerical values mentioned in the evaluation of the low limit condition \ref{eq:lim1} and \ref{eq:lim2}.

We can rewrite the condition in general way using (\ref{eq:LimitEstimaceGen}) as
\begin{align}
L_{experiment} \ll \frac{1}{2\pi}\tau c,\label{eq:estimgeneral}
\end{align}
which connects the linear size of the detector with duration of the pulse in the experiment.

\subsection{Movement of particles}
Again, we will investigate the behaviour of test particles in $x$ direction in the mode $+$ and $\times$ for both models, ablation model and piston model. We will use the geodesic equation Eq.~(82), which can be rewritten in a form of ellipse.

\subsection{The amplitudes for ablation model}
First, we will look at the mode $+$ for the wave vector in $x$ direction, which is given by relations Eq.~(85) of the geodesic equations Eq.~(82) in \cite{Kadlecova2015}.

For convenience, we will shift the start of the coordinates to $z=f_{2}$, then the coordinates of TT will be $x=y=0$ and $z=0$ and $t=\tau+O(h)$.
Without loosing any information, we perform a phase shift, $+\pi/2$, and get $h_{zz}(\tau)=A^{a}_{+}\sin\omega\tau$.  When $\tau=0$ then $h_{zz}=h_{\tilde{z}\tilde{z}}(\tau)\neq 0$ and in fact the function $b_{I}$ diverge, therefore we will investigate the behavior in small are around zero $0<\tau < \epsilon$ where $\epsilon$ is small number. Generally the amplitudes are non--zero for $0<\tau<\epsilon$, because of the correction terms with $b_{I}$.  

The semi-minor axes are
\begin{equation}
a[1 \pm A\sin\omega\tau],\label{eq:semiminoraxes}
\end{equation}
where $A=\frac{1}{2}A^{a}_{+}$ and
\begin{align}\label{eq:Ampl}
&A\equiv A|_{\tilde{x}_{A}^{j}=0}=-\frac{4}{r}\frac{G}{c^4}S\rho_{0}v_{r}^2\nonumber\\
&\times\left(3(-v_{r}\tau)(-\frac{2}{3}+e^{-b_{I}}(b^2_{I}+1))-2e^{-b_{I}}z_{L}(3+1/v_{r})\right),
\end{align}
the explicit form of $\frac{1}{2}h_{zz}^{TT}(\tau)|_{\tilde{x}_{A}^{j}=0}$ then is
\begin{align}\label{eq:Amplitu}
\frac{1}{2}&h_{zz}^{TT}(\tau)|_{\tilde{x}_{A}^{j}=0}=\-\frac{4}{r}\frac{G}{c^4\omega}S\rho_{0}v_{r}^2\sin\omega\tau\nonumber\\
&\times\left(3(-v_{r}\omega\tau)(-\frac{2}{3}+e^{-b_{I}}(b^2_{I}+1))-2e^{-b_{I}}z_{L}\omega(3+1/v_{r})\right),
\end{align}
where the function $b_{I}$ becomes
\begin{equation}
b_{I}=\frac{z_{L}+v_{r}\tau}{v_{r}\tau}.\label{eq:bIf2}
\end{equation}
The negativity of the amplitude just means that the change will happen in the transversal direction to the positive one.

For specific values, $a=b=1\,{\rm mm}=0.1\,{\rm cm}$,  $I_{L}=50\,[\rm PW/{\rm cm}^2]$ and  $R_{t}=15.144[{\rm kg}/{\rm m}^3]$  for Carbon. The velocity $v_{r}=4.884\times 10^5 {\rm [m/s]}$ and $b_{I}=0.2$ for time $t=10^{-9}$ s. And $\omega=2\pi c/\lambda=6.26\times10^{9}$ where $\lambda=0.3$m. Then we get the amplitude $A=-\frac{4}{r}\frac{G}{c^4}S\rho_{0}v^2_{r}=-1.483\times 10^{-34}$.
The effect of the GW on test particles does not produce ellipses but circles which grow with in time with distance between each circles for $\tau=\pi/2\omega$ from $\tau=0$, then back to one circle at $\tau=\pi/2\omega$. Then the circles grow equi--distantly with time for $\tau=3\pi/2\omega$. This effect of expansion of the test particles is definitely connected to the mass non conservation in the ablation model.

\begin{figure}
\centering
\subfigure[The test particles at $\tau=0.1$.]{%
\label{fig:Tau0}
\includegraphics[width=0.25\textwidth]{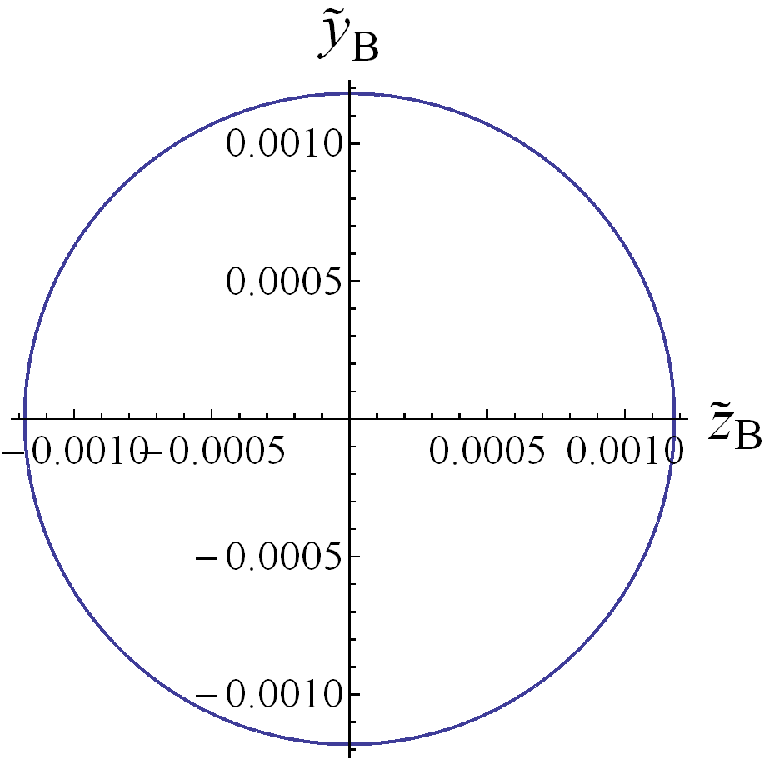}
}
\subfigure[The test particles at $\tau=\pi/2\omega; 5\pi/2\omega; 9\pi/2\omega; 13\pi/2\omega$ and more.]{%
\label{fig:Tau1}
\includegraphics[width=0.3\textwidth]{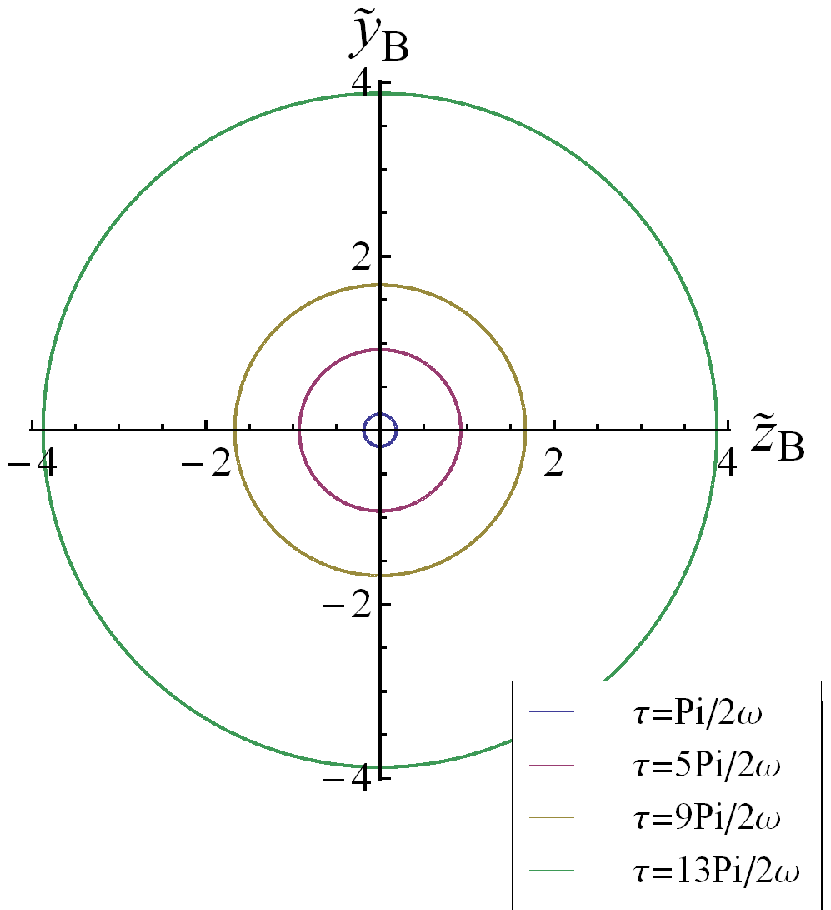}
}\\%
\subfigure[The test particles at $\tau=\pi/\omega;3\pi/\omega; 5\pi/\omega$ and more.]{%
\label{fig:Tau2}
\includegraphics[width=0.3\textwidth]{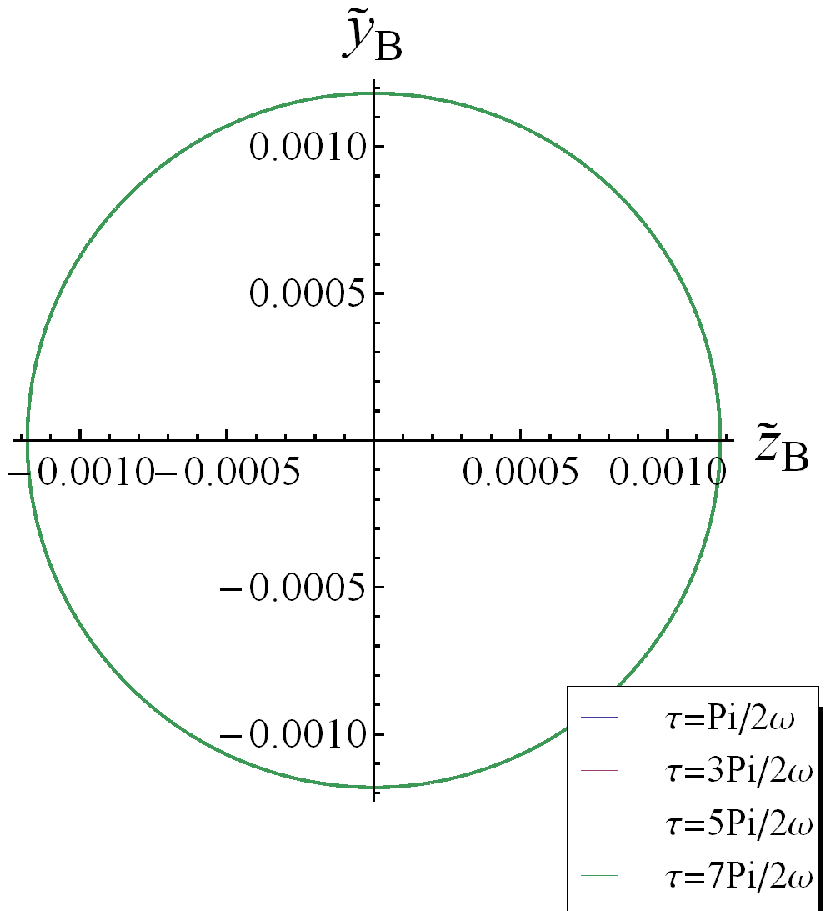}
}z
\subfigure[The test particles at $\tau=3\pi/2\omega; 7\pi/2\omega; 11\pi/2\omega$,$15\pi/2\omega$ and more.]{%
\label{fig:Tau3}
\includegraphics[width=0.3\textwidth]{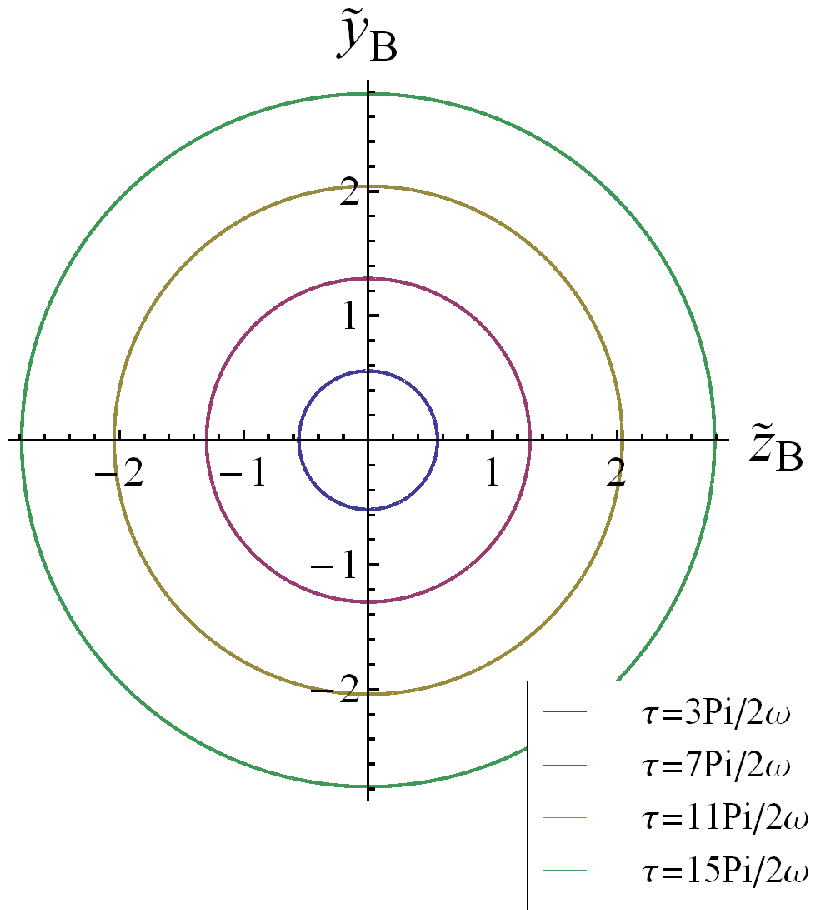}
}%
\caption{\label{fig:TestParticles} The diagrams depict the position of test particles in time evolution under influence of GW wave with $+$ polarization.}
\end{figure}

In the mode $\times$ we will get  deformation of a circle with the only non--zero component $h^{TT}_{zy}$. The equations of motion have form Eq.~(87) in \cite{Kadlecova2015} where the images for $+$ mode will be rotated for $45^{\circ}$.  Again, we perform a phase shift, $\pi/2$, and get $h_{yz}(\tau)=A^{a}_{\times}\sin\omega\tau$, then for $\tau=0$ we get $h_{yz}\neq 0$, the explicit form of $h_{yz}$ then become
\begin{align}\label{eq:Ampli}
\frac{1}{2}h_{yz}^{TT}(\tau)|_{\tilde{x}_{A}^{j}=0}=\frac{2}{r}\frac{G}{c^4}Sb\rho_{0}[2v_{r}^2(\frac{1}{2}+b_{I}e^{-b_{I}})]\sin{\omega\tau}.
\end{align}

\begin{figure}
\centering
\subfigure[The test particles at $\tau=0.01$, $\tau=2 \pi/\omega$ and more.]{%
\label{fig:Tau0}
\includegraphics[width=0.3\textwidth]{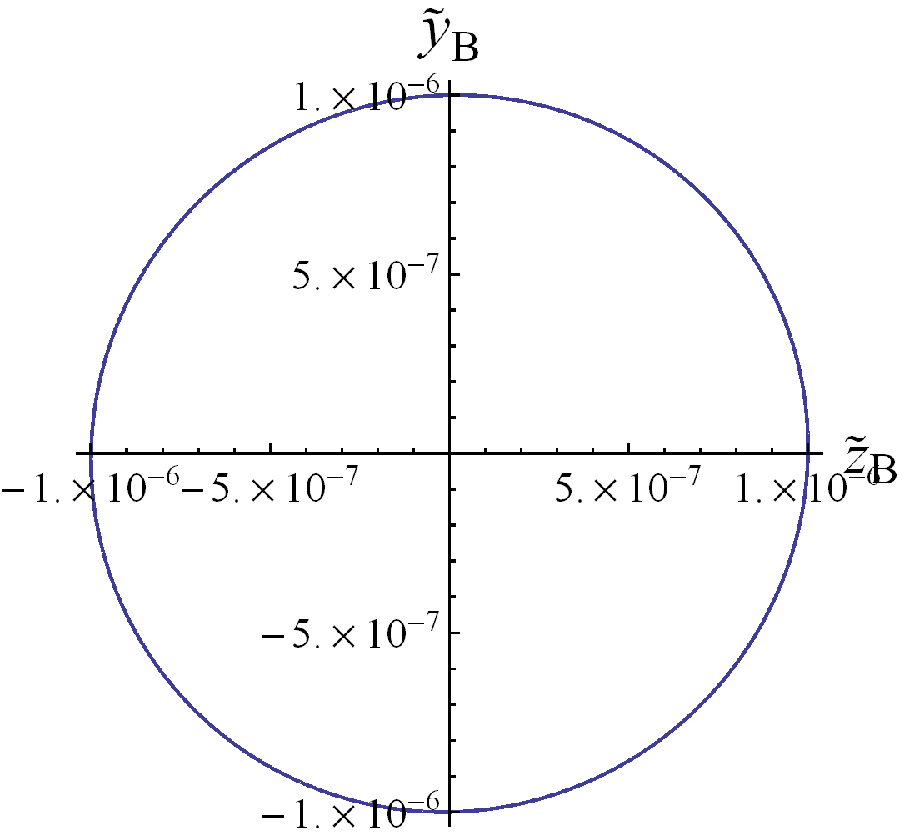}
}
\subfigure[The test particles at $\tau=\pi/2\omega; 5\pi/2\omega; 9\pi/2\omega; 13\pi/2\omega$ and more.]{%
\label{fig:Tau1}
\includegraphics[width=0.25\textwidth]{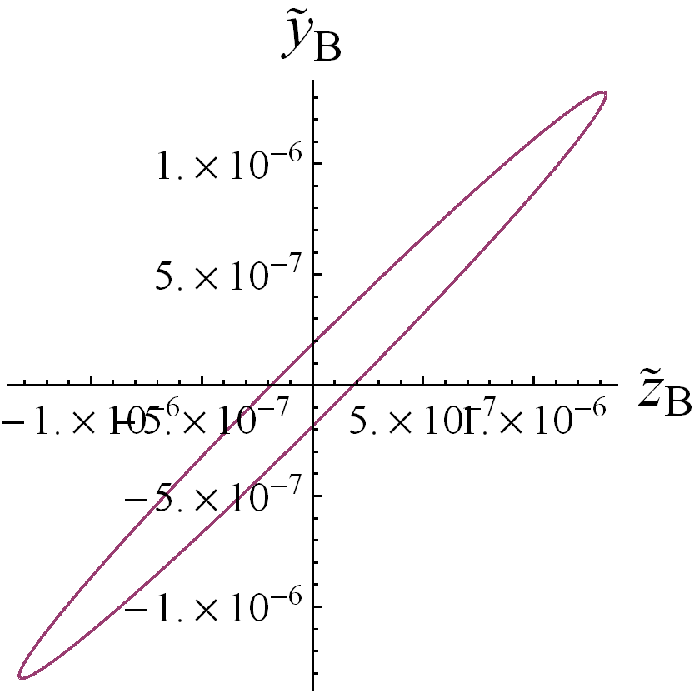}
}\\%
\subfigure[The test particles at $\tau=\pi/\omega;3\pi/\omega; 5\pi/\omega$ and more.]{%
\label{fig:Tau2}
\includegraphics[width=0.25\textwidth]{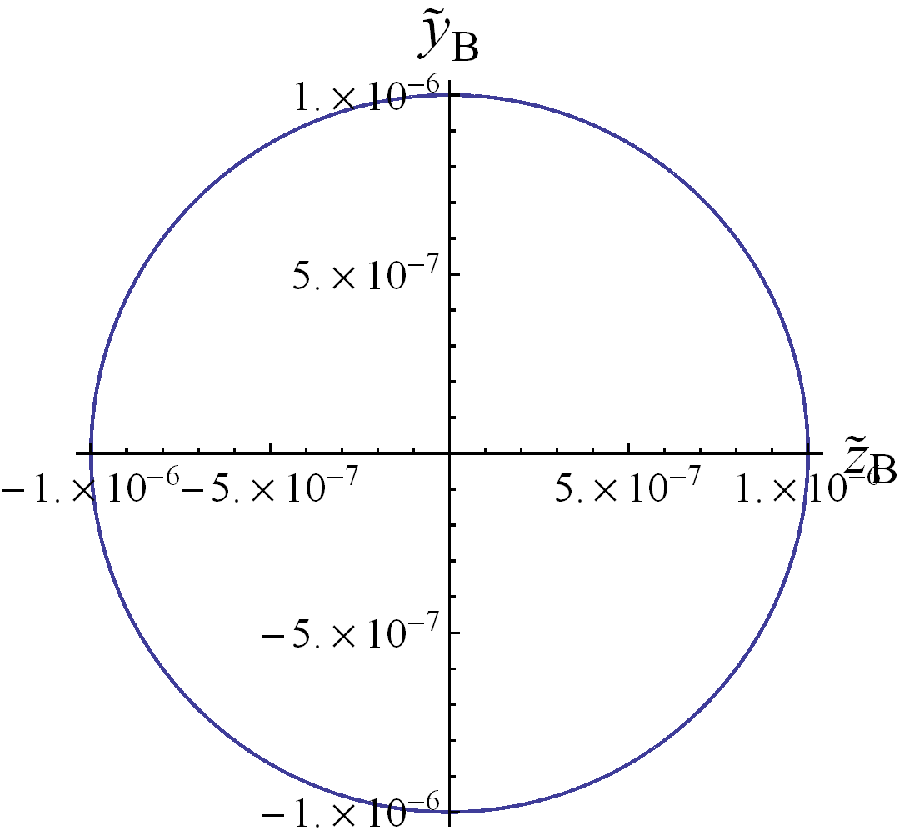}
}z
\subfigure[The test particles at $\tau=3\pi/2\omega; 7\pi/2\omega; 11\pi/2\omega$,$15\pi/2\omega$ and more.]{%
\label{fig:Tau3}
\includegraphics[width=0.25\textwidth]{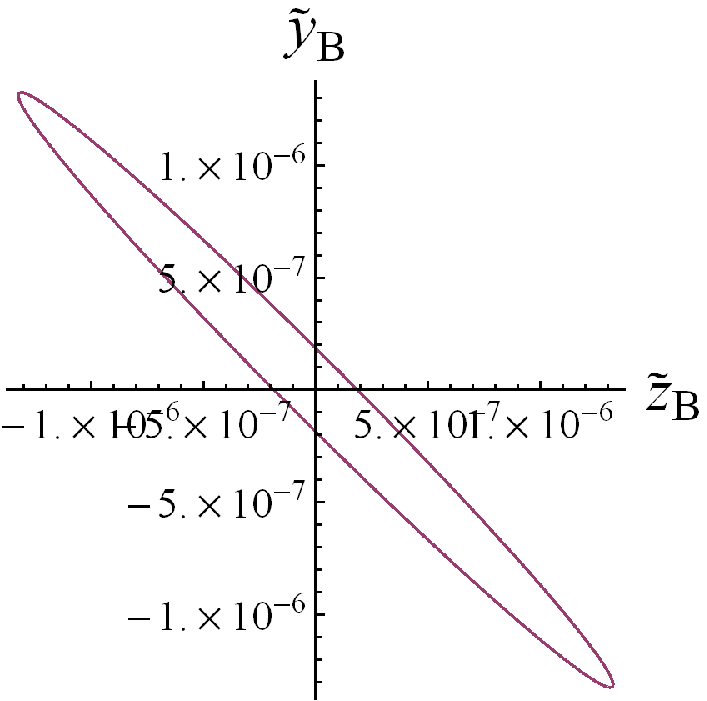}
}%
\caption{\label{fig:TestParticles} The diagrams depict the position of test particles in time evolution under influence of GW wave with $\times$ polarization.}
\end{figure}
The circle of test particles under influence of GW in mode $\times$ changes to the shapr ellipse of the same magnitude as the original circle at $\tau=0.01$. When we compare the images for this mode with shock wave model, we observe that the main difference is the much sharper shape of the ellipse.  

\subsection{The amplitudes for piston model}
In the mode $+$, is described by the Eq.~(85) in \cite{Kadlecova2015} and the semi-minor axes (\ref{eq:semiminoraxes}) and the $A=\frac{1}{2}A^{p}_{+}$ are for piston model
\begin{align}\label{eq:AmplPiston}
A\equiv A|_{\tilde{x}_{A}^{j}=0}=-\frac{1}{r}\frac{G}{c^4}S\rho_{0}v_{p}^3\tau,
\end{align}
and the explicit form of $\frac{1}{2}h_{zz}^{TT}(\tau)|_{\tilde{x}_{A}^{j}=0}$ then is
\begin{align}\label{eq:AmpliPiston}
\frac{1}{2}h_{zz}^{TT}(\tau)|_{\tilde{x}_{A}^{j}=0}=-\frac{1}{r}\frac{G}{c^4\omega}S\rho_{0}v_{p}^2(v_{p}\omega\tau)\sin{\omega\tau}.
\end{align}
The negativity of the amplitude just means that the change will happen in the transversal direction to the positive one.

In the mode $\times$ we will get also deformation of a circle with the only non--zero component $h^{TT}_{zy}$, that is zero for $\tau=0$, according to Eq.~(87) in \cite{Kadlecova2015}. The component $h^{TT}_{yz}$ then becomes
\begin{align}\label{eq:Ampli}
\frac{1}{2}h_{yz}^{TT}(\tau)|_{\tilde{x}_{A}^{j}=0}=\frac{1}{2r}\frac{G}{c^4}S\rho_{0}b v_{p}^2\sin{\omega\tau}.
\end{align}

The component $h_{yz}^{TT}$ has constant amplitude therefore the ellipses do not change shape when time grows and the images for piston model will appear the same as for the shock wave model Fig.~(11) and (12) in \cite{Kadlecova2015}.

In this section, we have investigated behaviour of test particles in the presence of GW with two modes of polarization for ablation and piston models. 

The main result of this section is that the time dependent amplitudes of polarization $+$ and $\times$ of ablation model influence the circle of particles to change the shape to larger circles in magnitude, at $\tau=\pi/2\omega$, and equidistant circles at $\tau=3\pi/2\omega$. In piston model, just the $+$ amplitude is time dependent and shapes the circle contrary to the $\times$ polarization which does not change the circle of test particles and the shape stays constant in time just for the piston model. This might serve as a measurable quality in the future experiments.

\section{\label{sec:number6}The conclusion}

In the second part of the paper, we have investigated the ablation and piston models for generation of gravitational waves for the possible experiments.

The ablation and piston models were investigated in linearized gravity  in quadrupole approximation which proved to be valid for the low velocity condition of the suggested experiments.

We have calculated and analyzed the perturbation tenzor $h^{TT}_{ij}$ and the luminosity of gravitational radiation ${\mathcal{L}}_{GW}$ in linear gravity in low (non-relativistic) velocity approximation far away from the source. We have generalized the results presented in \cite{grossmannMeet2009l,izestELINP2014} where we included the dependence on the laser wavelength and material of the foil for the ablation model. The calculations are presented in detail and estimations for real experimental values are included.
The ablation model has estimations for luminosity $\mathcal{L}=3.61\times 10^{-20}\,[{\rm erg/s}]$ and perturbation $h^{GW}_{zz}=4.7\times 10^{-39}$ for intensity $I_{L}=50\,[{\rm PW/cm^2}]$ and duration of pulse $1 {\rm ns}$. The piston model has luminosity $\mathcal{L}=2.7\times 10^{-18}\,[{\rm erg/s}]$ and perturbation $h^{GW}_{zz}=3\times 10^{-43}$ for intensity $I_{L}=7\times 10^{8}\,[{\rm PW/cm^2}]$ and duration of pulse $1 {\rm ps}$. Let us repeat that the luminosity for 
$\mathcal{L}=1.69\times 10^{-23}\,[{\rm erg/s}]$ and perturbation $h^{GW}_{zz}=2.37 \times 10^{-39}$ for intensity $I_{L}=0.5\times 10^{8}\,[{\rm PW/cm^2}]$ and duration of pulse $1\,{\rm ns}$, \cite{Kadlecova2015}.

The ablation model shows to have the highest luminosity of all the models and the perturbation of the same order as the shock wave model. Therefore the model might be the most suitable for the real experiment. In reality, it would depend on the technical realization of the possible model and the expenses.

Furthermore, we have investigated the two independent polarization modes of the gravitational radiation in the ablation and piston model. We have derived the amplitudes of the radiation in the three main directions of wave propagation, $x, y, z$. The radiation vanishes in the direction of motion in the $z$ direction for the piston model, the radiation in $\times$ mode appears for ablation model due to the fact that the model does not satisfy the mass conservation law and the existing radiation is a an artefact which vanishes in time and distance.

The radiation is non--vanishing in other directions as $x$ and $y$ directions, the amplitude for mode $+$ of the polarization occurs is time dependent and the other amplitude for mode $\times$ is time--indepenent for piston model. For ablation model, the amplitudes are both time--dependent. This fact might be measured in the real experiment.

We have also investigated the amplitudes in the general wave direction given by angles $\theta$ and $\phi$. Again the amplitudes are for both modes time--dependent in case of ablation model and for piston model the mode $+$ is time dependent and the mode $\times$ is time--independent in the general case. The result might be used for convenient positioning of detectors in real experiment.
The $+$ amplitude have toroidal symetry around $z=0$ axes for both ablation, piston and shock wave models. For ablation model, the $+$ amplitude is decreasing in magnitude with the distance as in shock wave model, while for piston model, the amplitude is slowly increasing in the magnitude with the distance from the source.
The $\times$ amplitude has a shape of a ball which has one point attached to the $z=0$ aches and remains constant in time and has much smaller amplitude than the $+$ amplitude.

The general directional structure of the radiation produces by the models has toroidal shape with symetry around $z$ axes for both models, the structure of ablation model has additional radiation along the $z$ axes which is caused by the model does not satisfy the mass conservation and non--zero radiation appears as its consequence. The radiation vanishes as the distance approches infinity.
The angular momentum for all models is vanishing due to the one dimensional character of the models.

Moreover, we have analyzed the influence of gravitational waves on test particles thanks to the geodesics equation. The effects of GW on test particles for piston models are similar to shock wave model \cite{Kadlecova2015} where the time--dependent amplitudes changes shape of the ellipse in time contrary to the constant amplitude $\times$ which does not change the shape of the ellipse. In ablation model, both amplitudes $+$ and $\times$ are time--dependent and the $+$ mode amplitudes shape changes just in magnitude as the time progresses, the change to larger circles is growing for $\tau=\pi/2\omega$ which change back to circle for $\tau=\pi/2$ and then they change to circles at higher magnitude which are equidistant for $\tau=3\pi/2/omega$ and its higher periods. The $\times$ mode changes the circle to sharp ellipse and back to circle as the shock wave model, but the ellipse for ablation model is much sharper.
All of the analyzed aspects of the GW radiation might be used to set up the possible experiment in the future.

The remaining problem of the models is the detection of the gravitational waves which have the amplitude of the metric perturbation around $10^{-40}$.

\section*{Acknowledgments}
H. Kadlecov\'{a} wishes to thank Tom\'{a}\v{s} Pech\'{a}\v{c}ek for many valuable discussions and reading the manuscript.
The work is supported by the project ELI - Extreme Light Infrastructure – phase 2 (CZ$.02.1.01/0.0/0.0/15\_008/0000162$ ) from European Regional Development Fund.

\appendix

\appendix
\section{\label{App:A1}The derivatives of an ansatz for $z_{s}$}
The derivatives of the arbitrary function $z_{i}(t)$ are 
\begin{align}
(z^2_{i})^{\dot{}}&=2z_{i}\dot{z}_{i},\quad (z^2_i)^{\ddot{}}=2\left\{(\dot{z}_{i})^2+z_{i}(z_{i})^{\ddot{}}\right\},\nonumber\\
(z^2_i)^{\dddot{}}&=2\left\{3\dot{z}_{i}\ddot{z}_{i}+z_{i}(z_{i})^{\dddot{}}\right\},\nonumber\\
(z^3_i)^{\dot{}}&=3z_{i}^2\dot{z}_{i},\quad (z^3_i)^{\ddot{}}=3z_{i}\left\{2(\dot{z}_{i})^2+z_{i}\ddot{z}_{i}\right\},\nonumber\\
 (z^3_i)^{\dddot{}}&=3\left\{2(\dot{z}_{i})^3+6z_{i}\dot{z}_{i}\ddot{z}_{i}+z_{i}^2(z_{i})^{\dddot{}}\right\}.\label{eq:derivativesOfZgen}
\end{align}
and after the ansatz for the $z_{r}$ (Ablation) we get
\begin{align}
{z}_{r}&=-v_{r}t+d,\quad\dot{z}_{r}=-v_{r},\quad \ddot{z}_{r}=0,\;
(z^2_{r})^{\dot{}}=2(-v_{r}t+d)\dot{z}_{r},\nonumber\\
(z^2_r)^{\ddot{}}&=2v^2_{r},\quad (z^2_r)^{\dddot{}}=0,\nonumber\\
(z^3_r)^{\dot{}}&=-3v_{r}(-v_{r}t+d)^2,\quad (z^3_r)^{\ddot{}}=6v^2_r(-v_{r}t+d),\nonumber\\
(z^3_r)^{\dddot{}}&=-6v_{r}^3,
(z^3_r)^{\ddddot{}}=0.\label{eq:derivativesOfZ}
\end{align}
and for $z_{p}$ (Piston) we get
\begin{align}
{z}_{p}&=v_{p}t,\quad\dot{z}_{p}=v_{p},\quad \ddot{z}_{p}=0,\nonumber\\
(z^2_{p})^{\dot{}}&=2v^2_{p}t,\quad (z^2_p)^{\ddot{}}=2v^2_{p},\quad (z^2_p)^{\dddot{}}=0,\quad
(z^3_p)^{\dot{}}&=3v_{p}^3t^2,\nonumber\\
(z^3_p)^{\ddot{}}&=6v^3_{p}t,\quad (z^3_p)^{\dddot{}}=6v_{p}^3.\label{eq:derivativesOfZpiston}
\end{align}

\section{Integrals for the ablation model}

The mass moment Eq.~(11) in \cite{Kadlecova2015} diagonal components then read
\begin{align}
M_{xx}&=\frac{4}{3}Sa^2\rho_{0}\left[z_{r}+\int_{z_{r}}^{z_{L}}e^{-m(z,t)} {\rm d} z\right],\nonumber\\
M_{yy}&=\frac{4}{3}Sb^2\rho_{0}\left[z_{r}+\int_{z_{r}}^{z_{L}}e^{-m(z,t)} {\rm d} z\right],\nonumber\\
M_{zz}&=4S\rho_{0}\left[z_{s}^3/2+\int_{z_{r}}^{z_{L}}z^2e^{-m(z,t)} {\rm d} z\right],\label{eq:componentsMassMomentDiagApp}
\end{align}
and non--diagonal components $M_{xy}, M_{yz}, M_{xz}$,
\begin{align}
M_{xy}&=S^2\rho_{0}\left[z_{r}+\int_{z_{r}}^{z_{L}}e^{-m(z,t)} {\rm d} z\right],\nonumber\\ 
M_{yz}&=2Sb\rho_{0}\left[z_{r}^2/2+\int_{z_{r}}^{z_{L}}ze^{-m(z,t)} {\rm d} z\right],\nonumber\\
M_{xz}&=2Sa\rho_{0}\left[z_{r}^2/2+\int_{z_{r}}^{z_{L}}ze^{-m(z,t)} {\rm d}z\right].
\label{eq:componentsMassMomentOffApp}
\end{align}

The integrals evaluate as 
\begin{align}
&\int_{z_{r}}^{z_{L}}e^{-m(z,t)} {\rm d} z=z_{r}(e^{-b_{I}}-e^{-a_{I}}),\nonumber\\
&\int_{z_{r}}^{z_{L}}ze^{-m(z,t)} {\rm d} z=z^2_{r}(b_{I}e^{-b_{I}}-a_{I}e^{-a_{I}}),\nonumber\\
&\int_{z_{r}}^{z_{L}}z^2e^{-m(z,t)} {\rm d} z=z_{r}^3[(1+b_{I}^2)e^{-b_{I}}-(1+a_{I}^2)e^{-a_{I}}]\label{eq:integrals}.
\end{align}

\section{Derivatives for the piston model}

\subsection{\label{A2}The derivatives of mass moment and quadrupole moment with $z_{p}$ function}
For calculation purposes we will present derivatives, first, second and third derivatives with respect to time, of the quadrupole moments here. The first derivatives of non--diagonal components are
\begin{align}
\dot{I}_{xy}&=\dot{M}_{xy}=\frac{1}{4}S^2\rho_{0}\dot{z}_p,\nonumber\\
\dot{I}_{yz}&=\dot{M}_{yz}=\frac{1}{4}Sb\rho_{0}(z^2_p)^{\dot{}},\nonumber\\ 
\dot{I}_{xz}&=\dot{M}_{xz}=\frac{1}{4}Sa\rho_{0}(z^2_p)^{\dot{}},\label{eq:firstQuadrMomentOff}
\end{align}
and the second derivatives are
\begin{align}
\ddot{I}_{xy}&=\ddot{M}_{xy}=\frac{1}{4}S^2\rho_{0}\ddot{z}_p,\nonumber\\
\ddot{I}_{yz}&=\ddot{M}_{yz}=\frac{1}{4}Sb\rho_{0}(z^2_p)^{\ddot{}},\nonumber\\
\ddot{I}_{xz}&=\ddot{M}_{xz}=\frac{1}{4}Sa\rho_{0}(z^2_p)^{\ddot{}}\label{eq:secondQuadrMomentOff}
\end{align}
and the third derivatives are 
\begin{align}
\dddot{I}_{xy}&=\dddot{M}_{xy}=\frac{1}{4}S^2\rho_{0}\dddot{z}_p,\nonumber\\ \dddot{I}_{yz}&=\dddot{M}_{yz}=\frac{1}{4}Sb\rho_{0}(z^2_p)^{\dddot{}},\nonumber\\
\dddot{I}_{xz}&=\dddot{M}_{xz}=\frac{1}{4}Sa\rho_{0}(z^2_p)^{\dddot{}}.\label{eq:secondQuadrMomentOff}
\end{align}

The derivatives of diagonal componets of the mass moments  are
\begin{align}
\dot{M}_{xx}&=\frac{Sa^2}{3}\rho_{0}\dot{z}_{p}, \dot{M}_{yy}=\frac{Sb^2}{3}\rho_{0}\dot{z}_{p},
\dot{M}_{zz}=\frac{S\rho_{0}}{3}(z^3_{p})^{\dot{}},\nonumber\\
\ddot{M}_{xx}&=\frac{Sa^2}{3}\rho_{0}\ddot{z}_{p},\ddot{M}_{yy}=\frac{Sb^2}{3}\rho_{0}\ddot{z}_{p},
\ddot{M}_{zz}=\frac{S\rho_{0}}{3}(z^3_p)^{\ddot{}},\nonumber\\
\dddot{M}_{xx}&=\frac{Sa^2}{3}\rho_{0}\dddot{z}_{p},\dddot{M}_{yy}=\frac{Sb^2}{3}\rho_{0}\dddot{z}_{p},
\dddot{M}_{zz}=\frac{S\rho_{0}}{3}(z^3_p)^{\dddot{}}.\label{eq:secondMassMoment}
\end{align}
The derivatives of the trace of the mass moment,
\begin{align}
(Tr M)^{\dot{}}&=\frac{S\rho_{0}}{3}\left\{(a^2+b^2)\dot{z}_{s}+(z^3_s)^{\dot{}}\right\},\nonumber\\
(Tr M)^{\ddot{}}&=\frac{S\rho_{0}}{3}\left\{(a^2+b^2)\ddot{z}_{s}+(z^3_s)^{\ddot{}}\right\},\nonumber\\
(Tr M)^{\dddot{}}&=\frac{S\rho_{0}}{3}\left\{(a^2+b^2)\dddot{z}_{s}+({z^3}_s)^{\dddot{}}\right\}.\label{eq:AllDerTraceM}
\end{align}

The derivatives of diagonal components of the quadrupole moment are
\begin{align}
\dot{I}_{xx}&=\frac{1}{9}S\rho_{0}\left\{(2a^2-b^2)\dot{z}_{p}-(z^3_{p})^{\dot{}}\right\},\nonumber\\
\dot{I}_{yy}&=\frac{1}{9}S\rho_{0}\left\{(2b^2-a^2)\dot{z}_{p}-(z^3_{p})^{\dot{}}\right\},\nonumber\\
\dot{I}_{zz}&=\frac{1}{9}S\rho_{0}\left\{2(z^3_{p})^{\dot{}}-(a^2+b^2)\dot{z}_{p}\right\},\label{eq:firstDerQuadrDiag}
\end{align}
the second derivatives
\begin{align}
\ddot{I}_{xx}&=\frac{1}{9}S\rho_{0}\left\{(2a^2-b^2)\ddot{z}_{s}-(z^3_{s})^{\ddot{}}\right\},\nonumber\\
\ddot{I}_{yy}&=\frac{1}{9}S\rho_{0}\left\{(2b^2-a^2)\ddot{z}_{s}-(z^3_{s})^{\ddot{}}\right\},\nonumber\\
\ddot{I}_{zz}&=\frac{1}{9}S\rho_{0}\left\{2(z^3_{s})^{\ddot{}}-(a^2+b^2)\ddot{z}_{s}\right\},\label{eq:secondDerQuadrDiag}
\end{align}
and third
\begin{align}
\dddot{I}_{xx}&=\frac{1}{9}S\rho_{0}\left\{(2a^2-b^2)\dddot{z}_{s}-(z^3_{s})^{\dddot{}}\right\},\nonumber\\
\dddot{I}_{yy}&=\frac{1}{9}S\rho_{0}\left\{(2b^2-a^2)\dddot{z}_{s}-(z^3_{s})^{\dddot{}}\right\},\nonumber\\
\dddot{I}_{zz}&=\frac{1}{9}S\rho_{0}\left\{2(z^3_{s})^{\dddot{}}-(a^2+b^2)\dddot{z}_{s}\right\},\label{eq:secondDerQuadrDiag}
\end{align}
When using the ansatz for the function $z_{p}$ (\ref{eq:zp}) some derivatives simplify significantly. Let us mention that to this point, we did not use the ansatz for $z_{p}$ (\ref{eq:zp}) and the every formula was derived for general function of time $z_{p}(t)$.

\subsection{\label{A3}The derivatives of the mass moment and quadrupole moment with substitution for $z_{s}$}
The derivatives of the non-diagonal components of quadrupole moment read
\begin{align}
\dot{I}_{xy}&=\dot{M}_{xy}=\frac{1}{4}S^2\rho_{0}v_{p},\quad \ddot{I}_{xy}=\ddot{M}_{xy}=0,\nonumber\\
\dot{I}_{yz}&=\dot{M}_{yz}=\frac{1}{2}Sb\rho_{0}v^2_{p}t,\quad \ddot{I}_{yz}=\ddot{M}_{yz}=\frac{1}{2}Sb\rho_{0}v^2_{p}, \label{eq:firstMassSubsZ}\\ 
\quad \dot{I}_{xz}&=\dot{M}_{xz}=\frac{1}{2}Sa\rho_{0}v^2_{p}t,\quad \ddot{I}_{xz}=\ddot{M}_{xz}=\frac{1}{2}Sa\rho_{0}v^2_{p}, \nonumber\\
\dddot{I}_{xy}&=\dddot{M}_{xy}=0,\,\dddot{I}_{yz}=\dddot{M}_{yz}=0,\,
\dddot{I}_{xz}=\dddot{M}_{xz}=0 \nonumber
\end{align}
and diagonal components of the mass moment
\begin{align}
\dot{M}_{xx}&=\frac{Sa^2}{3}\rho_{0}v_{p},\quad \ddot{M}_{xx}=\dddot{M}_{xx}=0,\nonumber\\
\dot{M}_{yy}&=\frac{Sb^2}{3}\rho_{0}v_{p},\quad \ddot{M}_{yy}=\dddot{M}_{yy}=0, \label{eq:firstMassSubsZ}\\ 
\quad \dot{M}_{zz}&=\frac{S}{3}\rho_{0}v_{p}^3t^2,\quad \dddot{M}_{zz}=\frac{2S}{3}\rho_{0}v^3_{p}t, \;
\ddot{M}_{zz}=\frac{2S}{3}\rho_{0}v^3_{p}.\nonumber
\end{align}

The derivatives of diagonal components of the quadrupole moment are
\begin{align}
\dot{I}_{xx}&=\frac{1}{9}S\rho_{0}v_{p}\left\{(2a^2-b^2)-3v^2_{p}t^2\right\},\nonumber\\
\dot{I}_{yy}&=\frac{1}{9}S\rho_{0}v_{p}\left\{(2b^2-a^2)-3v^2_{p}t^2\right\},\nonumber\\
\dot{I}_{zz}&=\frac{1}{9}S\rho_{0}v_{p}\left\{6v^2_{p}t^2-(a^2+b^2)v_{p}\right\},\label{eq:firstDerQuadrDiag}
\end{align}
the second derivatives
\begin{align}
\ddot{I}_{xx}=-\frac{2S}{3}\rho_{0}v^3_{p}t,&\quad \ddot{I}_{yy}=-\frac{2S}{3}\rho_{0}v^3_{p}t,\quad
\ddot{I}_{zz}=\frac{4S}{3}\rho_{0}v^3_{p},\label{eq:secondDerZQuadr}
\end{align}
and third derivatives
\begin{align}
\dddot{I}_{xx}&=-\frac{2}{3}S\rho_{0}v^3_{p},\,
\dddot{I}_{yy}=-\frac{2S}{3}\rho_{0}v^3_{p},\,
\dddot{I}_{zz}=\frac{4S}{3}\rho_{0}v^3_{p}.\label{eq:thirdDerZQuadr}
\end{align}

\bibliography{apssampELI}

\end{document}